\documentclass[twocolumn]{aastex63} 
\usepackage{amsmath}
\usepackage{natbib}
\usepackage{float}
\usepackage{hyperref}

\newcommand{\ldl}{$\lambda/\Delta\lambda$}
\newcommand{\teff}{$T_\mathrm{eff}$}
\newcommand{\logg}{$\log{g}$}

\newcommand{\vsini}{$v\sin{i}$}
\newcommand{\kms}{km~s$^{-1}$}
\newcommand{\mper}
{0.7106156$\pm$0.0000002~day}
\newcommand{\mecc}
{0.0088$\pm$0.0017}
\newcommand{\mprv}{23.70$\pm$0.05~km~s$^{-1}$}
\newcommand{\msrv}{28.41$\pm$0.06~km~s$^{-1}$}
\newcommand{\msysrv}
{+6.61$\pm$0.02~km~s$^{-1}$}
\newcommand{\manom}{$1.98^{+0.14}_{-0.15}$~rad}
\newcommand{\marg}
{$1.20^{+0.15}_{-0.14}$~rad}
\newcommand{\mq}{0.8340$\pm$0.0017}
\newcommand{\ecosw}{$0.034^{+0.013}_{-0.014}$}
\newcommand{\esinw}{$0.086^{+0.009}_{-0.010}$}
\newcommand{\sigmaj}{$-2.48^{+0.13}_{-0.12}$~km~s$^{-1}$}
\newcommand{\masspri}{0.076--0.087} 
\newcommand{\masssec}{0.063--0.073}
\newcommand{\incshort}{24}
\newcommand{\inc}{24.3$\pm$0.6}
\newcommand{\radii}{0.093--0.109}
\newcommand{\separation}{0.0081--0.0084}
\newcommand{\sepradii}{17--19} 

\submitjournal{ApJ Letters}

\shorttitle{Short-Period Ultracool Binary LP 413-53}
\shortauthors{Hsu, Burgasser, \& Theissen}

\graphicspath{{./}{figures/}}

\begin{document}

\title{Discovery of the Exceptionally Short Period Ultracool Dwarf Binary LP 413-53AB}

\correspondingauthor{Chih-Chun Hsu}
\email{chsu@northwestern.edu}

\author[0000-0002-5370-7494]{Chih-Chun Hsu}
\affil{Center for Interdisciplinary Exploration and Research in Astrophysics (CIERA), Northwestern University,
1800 Sherman, Evanston, IL, 60201, USA}
\affil{Center for Astrophysics and Space Science, University of California San Diego, La Jolla, CA 92093, USA}

\author[0000-0002-6523-9536]{Adam J.\ Burgasser}
\affil{Center for Astrophysics and Space Science, University of California San Diego, La Jolla, CA 92093, USA}

\author[0000-0002-9807-5435]{Christopher A.\ Theissen}
\altaffiliation{NASA Sagan Fellow}
\affil{Center for Astrophysics and Space Science, University of California San Diego, La Jolla, CA 92093, USA}

\begin{abstract}

We report the detection of large-amplitude, rapid radial velocity (RV) variations and line-splitting in high-resolution Keck/NIRSPEC spectra of the M9 dwarf LP~413-53. 
We attribute these features to binary motion.
Analyzing data spanning \replaced{9~months}{15~years}, we
infer \replaced{an}{a preliminary} orbital period of {\mper},
an eccentricity of {\mecc},
a primary RV semi-amplitude of {\mprv},
and a secondary RV semi-amplitude of {\msrv}, 
implying a system mass ratio $M_\mathrm{secondary}$/$M_\mathrm{primary}$ = {\mq}.
These measurements identify LP~413-53 as the shortest-period
ultracool binary discovered to date, and one of the smallest separation
main sequence binaries known.
The position and velocity of the system rule\deleted{s} out previously reported membership 
in the Hyades Moving Group, 
and indicate that this is likely a pair of evolved (age $\gtrsim$ 1~Gyr), very-low-mass stars.
Assuming masses consistent with evolved late-M and L dwarfs, 
we estimate an orbital separation of {\separation}~au
or {\sepradii} stellar radii, and an orbital inclination angle of {\incshort}$\degr$,
making it unlikely that this system exhibits eclipse events.
The larger radii of these stars at young ages would have put them \replaced{in}{near} contact 
at the system's current separation, and we speculate that this system
has undergone dynamical evolution, either through orbital angular momentum loss or
ejection of a third component followed by tidal circularization. 
\added{While further observations are needed to fully constrain the orbital and physical parameters of LP~413-53, this ultra-short-period UCD binary system serves as a new testbed for formation and dynamical evolution models of very-low-mass multiples.}
\end{abstract}


\section{Introduction} \label{sec:intro}


Multiple systems are fundamental in studies of star formation, evolution, and physical characterization.
These systems \added{can} provide direct
measurements of stellar mass, 
while observed distributions of orbital separation, period, mass ratio, and eccentricity constrain star formation theory.
The closest-separation systems are particularly useful given their short orbital cadence and the opportunity to measure both masses and radii through mutual eclipses, 
essential for calibrating stellar evolution and structure models.

The frequency of stellar multiplicity is known to scale with system mass, with 
the binary fraction declining from 
$>$80\% for the most massive O-type stars to $<$20\% for the lowest-mass stars and brown dwarfs (\citealt{Duchene:2013aa} and references therein).
The latter ultracool dwarf (UCD) binaries ($M <$ 0.1~M$_{\odot}$)
also have smaller orbital separations and eccentricities, and higher secondary-to-primary mass ratios, compared to solar-type stars \citep{2007prpl.conf..427B,Dupuy:2017aa,2018MNRAS.479.2702F}. 
The majority of UCD binaries have been discovered as resolved systems, 
while the most closely-separated binaries have been 
identified as overluminous sources \citep{Smart:2019aa}, 
spectral blend binaries \citep{Burgasser:2010aa,Bardalez-Gagliuffi:2014aa},
and astrometric or radial velocity \added{(RV)} variables \citep{1999AJ....118.2460B,Blake:2010aa,Sahlmann:2020aa,Hsu:2021aa}.
Only three eclipsing UCD binaries have so far been reported in the literature: 
2MASS~J0535$-$0546AB \citep{Stassun:2006aa}, 
USCO~J1616$-$2512 \citep{Lodieu:2015aa}, and 
SPEC~J1510$-$2818AB \citep{Triaud:2020aa}, the second of these
having the shortest period at 2.8~days.
For comparison, the closest-separation (non-UCD) binaries have periods 
down to 0.2~day \citep{1997PASP..109..782M,2010MNRAS.406.2559D,2012MNRAS.425..950N,2013ApJ...764...62D,2019PASP..131c4201Z}.
The three known eclipsing UCD binaries are all members of young star forming regions and associations (1--40~Myr) and have inflated radii consistent with ongoing contraction.
No fully-evolved UCD eclipsing binary has been identified to date,
although several transiting UCD companions to more massive stars have been found
by missions such as the Transiting Exoplanet Survey Satellite (\textit{TESS}; \citealt{2016SPIE.9904E..2BR,2021AJ....161...97C}).

LP 413-53 (aka 2MASS~J03505737+1818069) is an M9 dwarf \citep{Gizis:2000aa} that has exhibited multiple signs of unresolved multiplicity in observations conducted over the past twenty years.
Its \textit{Gaia} Data Release 3 (DR3; \citealt{2022arXiv220800211G}) absolute $G$-band magnitude, $M_G$ = 14.376$\pm$0.010,
is nearly a full magnitude brighter than other M9 dwarfs (15.3$\pm$0.5; \citealt{2019AJ....157..231K}), while its \textit{Gaia}
Renormalized Unit Weight Error (RUWE) value of 1.35 is in the range where binary-induced astrometric variability is likely
\citep{2020MNRAS.496.1922B}. 
\citet{Reid:2002aa} obtained high-resolution optical spectroscopy of LP 413-53 
and reported a complex H$\alpha$ emission feature, consisting of a narrow spike on top of a broad peak. 
\replaced{There was also disagreement between t}{T}he RV inferred from the peak of the H$\alpha$ line ($5.8 \pm 2.0$~{\kms}) \replaced{and}{was inconsistent with} other spectral features ($-14.3 \pm 0.4$~{\kms}).
While it was noted that this behavior was similar to that of a known Hyades double-lined spectroscopic binary (SB2), the authors ultimately concluded that they had captured a flare during the observation.
\citet{Deshpande:2012aa} obtained high-resolution near-infrared spectroscopy of LP 413-53
and reported an RV of $32.2 \pm 1.8$~{\kms}, 
discrepant with both measurements from \citet{Reid:2002aa}.
In addition to the complex H$\alpha$ emission morphology reported by \citet{Reid:2002aa}, this source exhibits long-term emission variability, with equivalent widths ranging from undetectable ($<$1~{\AA}; \citealt{Gizis:2000aa})
to typical for a late-type M dwarf (13~{\AA}; \citealt{Cruz:2002aa})
to unusually high (40~{\AA}; \citealt{Reid:2002aa}). 
This variability could be an indicator of youth, and LP 413-53 has been reported as a possible kinematic member of the $\sim$650~Myr Hyades cluster \citep{Goldman:2013aa, Lodieu:2020aa}.
High angular resolution imaging observations obtained with the \textit{Hubble Space Telescope} \citep{Gizis:2003aa, Bouy:2003aa} and ground-based adaptive optics \citep{Close:2003aa} have failed to detect a companion
down to 0$\farcs$1 ($\sim$4~au), and no co-moving companion is present in \textit{Gaia} data out to 
 \added{a radius of} 1$\degr$.

In this Letter, we report new high-resolution spectra of LP~413-53 that show line-doubling and large-amplitude RV variations, identifying this source as the shortest-period UCD binary discovered to date.
In Section~\ref{sec:observe}, we describe our spectroscopic observations of LP 413-53.
In Section~\ref{sec:rv}, we determine RVs through forward modeling using a two-star model, which confirm the binary nature of the system.
In Section~\ref{sec:orbit}, we conduct orbit fits to our RV time series to determine orbital and physical parameters.
In Section~\ref{sec:discuss}, we discuss the properties of this system and comment on its formation and dynamical evolution.

\section{Observations} \label{sec:observe}
\subsection{Keck/NIRSPEC}
 
We obtained high-resolution near-infrared spectra of LP~413-53 using the Near-InfraRed SPECtrometer (NIRSPEC; \citealp{McLean:1998aa}) on the Keck II telescope. 
Data were obtained in \replaced{four}{five} UT epochs: 
2022 March 13, 2022 July 10, 2022 October 16, \replaced{and 2022 December 2}{2022 December 2, and 2023 January 8}. \replaced{all having}{The first four nights had} clear conditions with 0$\farcs$5--0$\farcs$6 seeing\added{, while the last night had 1$\farcs$0 seeing}.
In July, we used the NIRSPEC-7 filter and 0$\farcs$432 $\times$12$\arcsec$ slit
to obtain 1.84--2.63~$\micron$ spectra at a resolution of {\ldl} $\approx$ 35,000.
We obtained four exposures of 450~s each in two AB nodding pair sequences.
In the other epochs, we used the Kband-new filter, which covers 1.91--2.55~$\micron$ and provides 20\% higher throughput (James Lyke 2022, priv.\ comm.).
In March, we used the 0$\farcs$432$\times$12$\arcsec$ slit
and obtained six exposures of 300~s each in three AB sequences.
In October, we used the 0$\farcs$288$\times$12$\arcsec$ slit to acquire slightly higher resolution data ({\ldl} $\approx$ 52,000),
and obtained twelve exposures of 300~s each in four ABBA sequences; the last exposure was rejected due to high sky background during sunrise.
In December, we used the 0$\farcs$288$\times$12$\arcsec$ slit
and obtained twenty-one exposures of 300~s each in three ABBA sequences and an additional A nod position.
\added{In January, we used the 0$\farcs$432$\times$12$\arcsec$ slit
and obtained fifty-seven exposures of 300~s each in thirteen ABBA sequences, two AB sequences, and an additional B nod position.}
For all observations, we observed the A0~V stars HD~25175 or HD~174567
before or after our LP~413-53 exposures for telluric and wavelength calibration.
We also obtained internal dark frame, flat-field lamp, and arc lamp exposures at the beginning or end of the night for detector calibration.
Data were reduced using a modified version of the NIRSPEC Data Reduction Pipeline (NSDRP; \citealp{Tran:2016aa}), as detailed in \cite{Theissen:2022aa} and \cite{Hsu:2021aa}.
Median signals-to-noise (S/N) range from \replaced{8--30}{6-30} per pixel (Table~\ref{tab:nirspec}).

We also evaluated archival high-resolution NIRSPEC $J$-band spectra
available on the Keck Observatory Archive \citep{https://doi.org/10.26135/koa1},\footnote{\url{https://koa.ipac.caltech.edu/}.}
obtained on 2007 October 26-27 (UT) through program 2007B-N076NS (PI E.\ Mart\'{i}n) and reported in \cite{Deshpande:2012aa}.
We reanalyzed these data following the same procedures as our acquired data.

\section{Radial Velocity Measurements} \label{sec:rv}

\begin{figure*}[htbp!]
\centering
\includegraphics[width=0.9\textwidth]{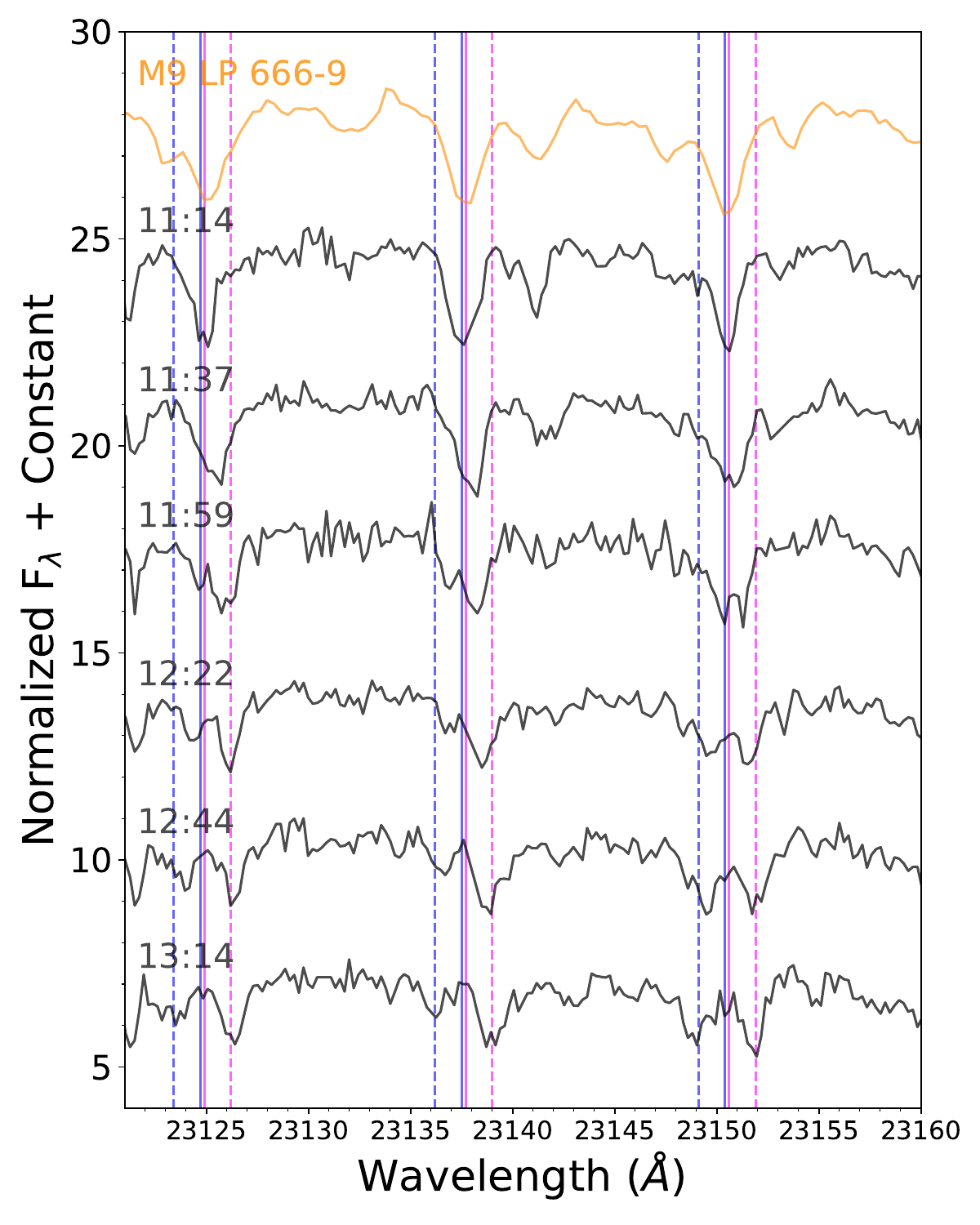}
\caption{Sequence of LP~413-53 spectra over a 2~hr period on 2022 December 2 (UT), centered on the 2.312--2.316~$\mu$m region. Time increases from top to bottom, and UT times (hour and minute) are indicated next to each spectrum. 
The topmost spectrum shows equivalent data for the M9 dwarf LP 666-9 (Hsu et al.\ in prep.).
Absorption features arising from CO are initially consistent with a single-line source, but become clearly doubled over the \replaced{observng}{observing} sequence, with the line centers of the stronger features drifting to the right and the line centers of the weaker features drifting to the left. Vertical lines mark line center positions at the start (solid) and end (dashed) of the sequence for the stronger (magenta) and weaker (blue) lines. 
}
\label{fig:raw_spec}
\end{figure*}

Figure \ref{fig:raw_spec} displays the sequence of NIRSPEC spectra of LP~413-53 from 2022 December 2, centered near $\lambda\sim2.31$~$\micron$, a region relatively free of telluric absorption.
The data are initially consistent with a singled-line spectrum, but over the course of the sequence we observe double-line splitting, with
the stronger line of each pair drifting to the right (positive velocity change) and the weaker line 
drifting to the left (negative velocity change). Similar behaviors are observed in \replaced{all}{the first} four epochs of our acquired NIRSPEC data\added{, while the last epoch showed the lines swapped and drifting in the opposite directions}.
Combined with the large variance of past RV measurements, we interpret these patterns as arising from the orbital motion of primary and secondary components in a close-separation binary.

To more precisely measure these trends in the NIRSPEC data, we used a Markov Chain Monte Carlo (MCMC) forward-modeling approach integrated in the \texttt{Spectral Modeling Analysis and RV Tool} (\texttt{SMART}) package \citep{Hsu:2021aa,Hsu:2021ab}.
This code fits individual ``raw'' (uncalibrated) high\replaced{ }{-}resolution near-infrared spectra to infer
RV, rotational velocity ({\vsini}), temperature ({\teff}), and surface gravity ({\logg}), as well as several calibration parameters.
For our acquired NIRSPEC observations, we focused our analysis on order 33 data
which covers the $^{12}$CO ($\nu$=2--0) bandhead (2.285--2.318~$\micron$) in the stellar spectra, and \added{the} telluric CH$_4$ \replaced{absorption}{$\nu_2$+$\nu_3$} Q branch\deleted{es} \citep{McMath:1949aa} at $\sim$2.315~$\micron$ which anchors the absolute wavelength calibration.
For the archival observations, we analyzed data in order 57, which covers wavelengths between 1.325--1.347~$\micron$.
We used BT-Settl atmosphere models \citep{Allard:2012aa} for the stellar components and ESO models for telluric absorption \citep{Moehler:2014aa}.
An initial fit assuming a single star model was made to obtain a first estimate of the four physical parameters of the primary listed above, as well as the telluric airmass (AM) and precipitable water vapor (\replaced{PMV}{PWV}). 
We then adapted the \texttt{SMART} code to implement a two-star model (cf.\ \citealt{Triaud:2020aa}), with independent values of 
the physical parameters of each component (8 stellar parameters), their relative flux scaling $f$ (1 parameter), the telluric model (2 parameters), and nuisance parameters (3 parameters\added{; flux, wavelength, and noise scale factor; see \citealt{Hsu:2021aa} for details}).

MCMC \replaced{modelling}{modeling} of each spectrum was done in two stages: an initial fit of the data was made to obtain preliminary parameters; a second fit of the data was then made after masking 2.5$\sigma$ flux outliers, which typically removes bad pixels.  
Each MCMC run was conducted with 50 chains and 1000 steps, with burn-ins of 600 and 500 steps for the initial and final fits, respectively. 
The best-fit parameters and associated uncertainties, determined from the 16\%, 50\%, and 84\% quantiles of the chains after removing burn-in,
are summarized in Table~\ref{tab:nirspec}.

\begin{figure*}[tbp!]
\centering
\includegraphics[width=1.0\textwidth]{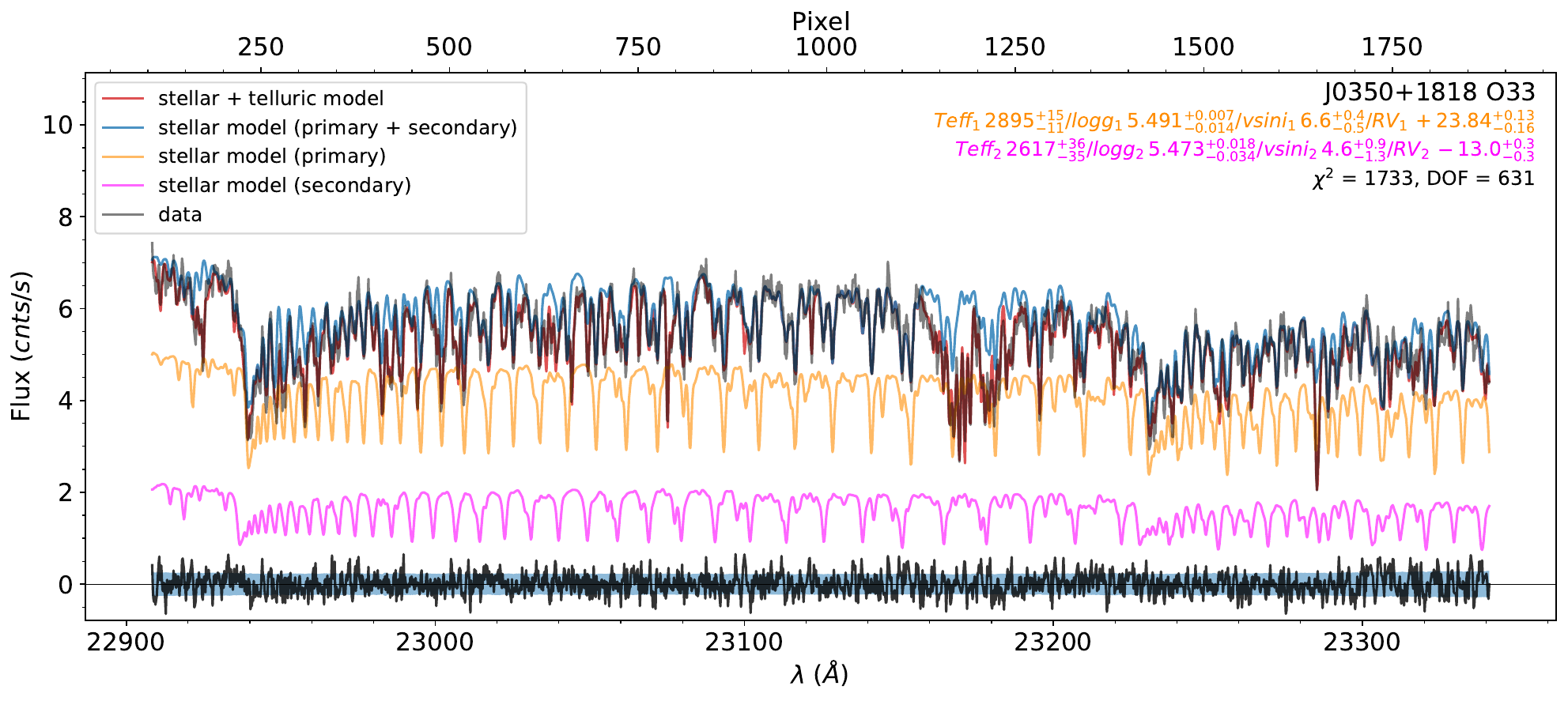}
\caption{
Forward-model fit to the 2.29--2.33~$\mu$m (order 33) Keck/NIRSPEC spectrum of LP~413-53 obtained on 2022 March 13 at 05:02:59.43 UT (\replaced{59651.21041}{59651.20898} MJD; black line).
The forward model consists of primary (orange) and secondary (magenta) spectra, with the combined spectrum (blue) modulated by telluric absorption (red).
The residual (data $-$ model) and uncertainty are depicted by the black line and grey-shaded region along the bottom, respectively.
The legend provides the final fit parameters for component temperatures ({\teff} in K), log surface gravities ({\logg} in log cm/s$^2$), rotational velocities ({\vsini} in {\kms}), and radial velocities (RV in {\kms}) with their uncertainties, 
as well as the best\replaced{ }{-}fit $\chi^2$ and degrees of freedom (DOF).
}
\label{fig:nirspec}
\end{figure*}

Figure~\ref{fig:nirspec} illustrates the best-fit model for data obtained on 2022 March 13 at 05:02:59.43 UT\added{, or} \deleted{(}\replaced{59651.21041}{59651.20898} MJD\added{\footnote{All modified Julian Dates (MJD) had the Barycentric Dynamical Time correction is applied following \cite{Eastman:2010aa} using the \texttt{JDUTC{\textunderscore}to{\textunderscore}BJDTDB} function in the \texttt{barycorrpy} package \citep{Kanodia:2018aa}.}}\deleted{)}.
The two-component model provides an excellent fit to the NIRSPEC spectrum, with flat residuals and a reduced $\chi_r^2 = 2.7$ significantly better than the equivalent single-star model ($\chi_r^2 = 7.6$). 
The component RVs in this epoch differ by 36.9$\pm$0.3~{\kms}, more than three times the spectral resolution\deleted{ at this epoch}.
Averaging over the several epochs of 2.3~$\micron$ measurements (Table~\ref{tab:orbit}),
we find that the weaker lines emerge from a component that is $f_{2.3}$= \replaced{0.60 $\pm$ 0.04}{0.65 $\pm$ 0.08} times as bright and \replaced{240}{250}~K cooler than the stronger line component, consistent with the expected ordering of secondary and primary.
The model temperature for the primary, {\teff} = \replaced{2877$\pm$41}{2879$\pm$37}~K, is somewhat warmer than the typical temperatures of M9 dwarfs (2370--2480~K; \citealt{2009ApJ...702..154S,Filippazzo:2015aa,Dupuy:2017aa}),
and likely reflects model bias common to fits of spectral data over narrow wavelength ranges (cf.\ \citealt{2010ApJS..186...63R}).
Both components have similar {\logg} $\approx$ 5.5 (cgs), consistent with evolved UCDs.
Both components also have similar {\vsini}s that are close to the resolution limit of the data \citep{Hsu:2021aa}.
These {\vsini} values are similar to the {\vsini} = 4~{\kms} reported in \citet{Reid:2002aa}, but significantly smaller than the 12.2$\pm$3.0~{\kms} reported in \citet{Deshpande:2012aa}, which could be explained by line doubling.
These results hold across all epochs, 
\deleted{. In particular, we find the binary star model is a consistently superior fit to a single star model} with component RV differences as high as \replaced{60}{50}~{\kms}.


\section{Radial Velocity Time Series and Binary Orbital Fit}\label{sec:orbit}

We measured the orbital parameters of LP 413-53 by fitting the primary and secondary RV measurements with \added{the archival NIRSPEC $J$-band and} our \added{acquired} NIRSPEC $K$-band data.
We used a two-component radial velocity orbit model
to measure the orbital parameters of the RV time series:
\begin{equation}
RV_1  = K_1 [ e\cos{\omega} + \cos{(T(t)+ \omega)} ] + v_0
\end{equation}
\begin{equation}
\begin{split}
RV_2  & = - K_2 \, [ e\cos{\omega} + \cos{(T(t)+ \omega)} ] + v_0   \\
& = - \frac{K_1}{q} [ e\cos{\omega} + \cos{(T(t)+ \omega)} ]  + v_0.
\end{split}
\end{equation}
Here, $RV_{1,2}$ and $K_{1,2}$ are the radial velocities and RV semi-amplitudes of the primary and secondary components, respectively; $q = M_2/M_1$ is the binary mass ratio; $e$ is the orbital eccentricity;
$\omega$ is the argument of periastron; $v_0$ is the systematic velocity; and $T(t)$ is the true anomaly of the orbit.
The mean anomaly, $M = M_0 + 2 \pi (t - t_0)/P $ with $t_0$ = \replaced{59868.60714~MJD}{59952.24622~MJD} and $P$ being the orbital period, is related to the true anomaly $T(t)$ through Kepler's equation, which we computed using the \texttt{kepler.py}\footnote{\url{https://github.com/dfm/kepler.py}} code.
\added{We also introduced an RV offset term for the $J$-band data relative to the $K$-band data,
$\sigma_{J}$, 
to account for interband calibration differences, a common approach for analyzing multi-instrument RV time series
(cf. \citealp{Howard:2014aa}).}
To determine the best-fit orbital parameters and their uncertainties, we used a 
Markov Chain Monte Carlo algorithm (MCMC) through the \texttt{emcee} package \citep{Foreman-Mackey:2013aa}.
Our log likelihood function $\mathcal{L}$ was defined as 
\begin{equation}
\ln{\mathcal{L}} = -0.5 \times (\chi_1^2 + \chi_2^2 + \Sigma_i{[\ln{2\pi \sigma_{1,i}^2}]} + \Sigma_i{[\ln{2\pi \sigma_{2,i}^2}]}),
\end{equation}
where $\sigma_{1,2}$ are the RV measurement uncertainties and $\chi_{1,2}^2 = \Sigma_i[(RV_\mathrm{measure,i} - RV_\mathrm{model,i}) / \sigma_i]^2$ for the primary and secondary components, respectively.
We chose a basis set of $P$, $K_1$, $K_2$, $\sqrt{e}\sin\omega$, $\sqrt{e}\cos\omega$, $v_0$, \deleted{and }$M_0$\added{, and $\sigma_{J}$} to simultaneously fit the RV time series of both components, following \citet{2005AJ....129.1706F}.
We \replaced{discarded}{excluded from our fit} the first four RV measurements on 2022 Dec 2\added{, and the three epochs of 59952.43355, 59952.43702, and 59952.44147 MJD on 2023 Jan 8,} \deleted{in our fit} as the unresolved RVs \added{at these epochs} are likely less reliable, \replaced{but}{although} our result is \added{largely} consistent with these measurements.
\deleted{We also excluded the $J$-band RV measurements \replaced{in}{from} the fit, as discussed below.}
We used 100 chains of \replaced{100,000}{400,000} steps each, with the first \replaced{50}{25}\% of each chain discarded as burn-in.
The chains converged in the first few hundred steps, with an integrated autocorrelation scale of \replaced{790}{370} steps.

\begin{figure*}[htbp!]
\centering
\added{\includegraphics[width=0.32\textwidth]{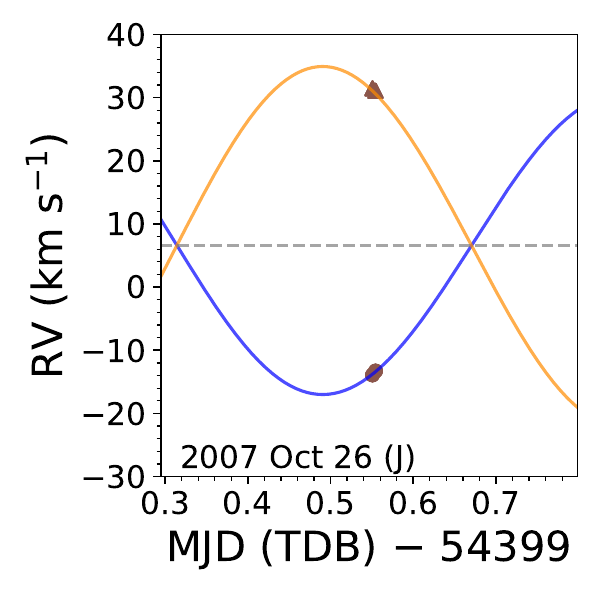}}
\added{\includegraphics[width=0.32\textwidth]{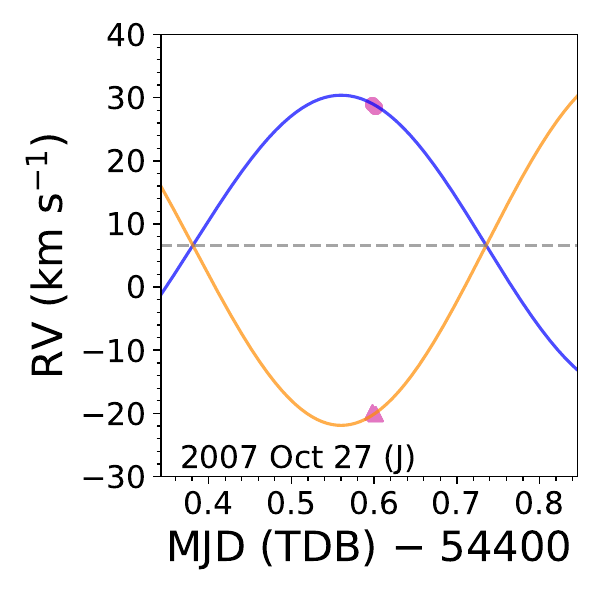}}
\includegraphics[width=0.32\textwidth]{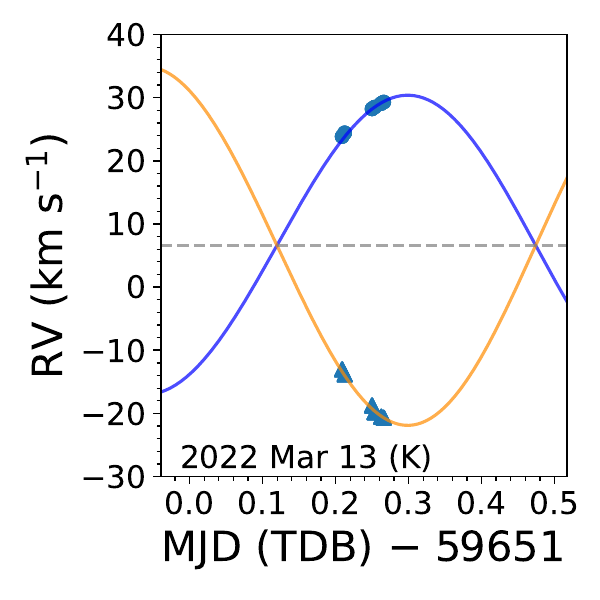} \\
\includegraphics[width=0.32\textwidth]{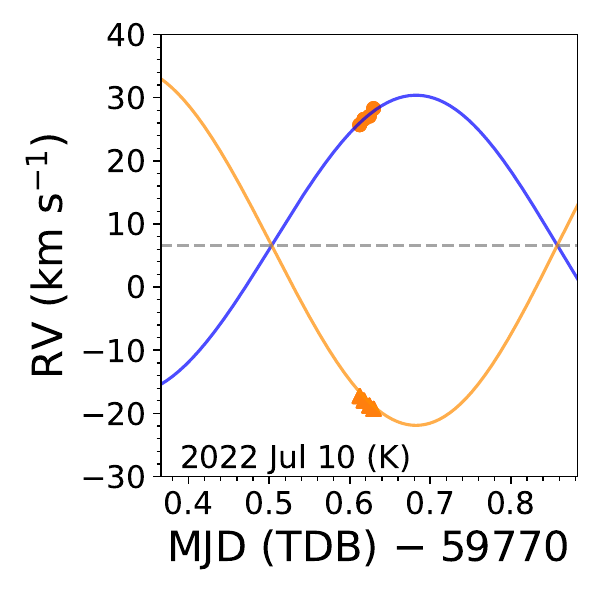}
\includegraphics[width=0.32\textwidth]{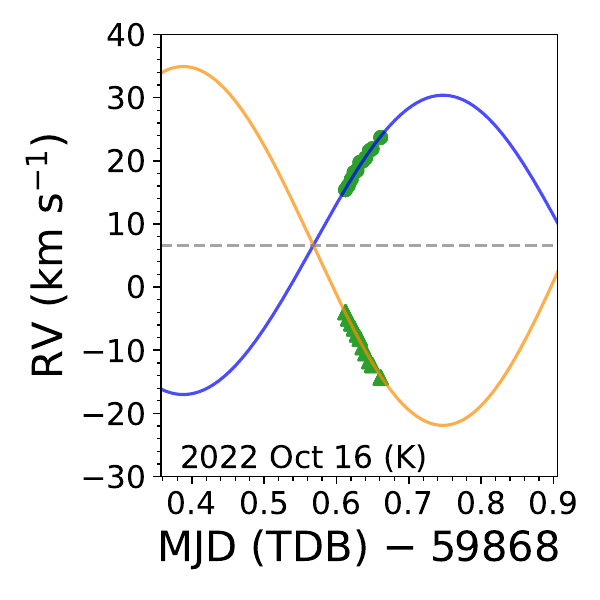}
\includegraphics[width=0.32\textwidth]{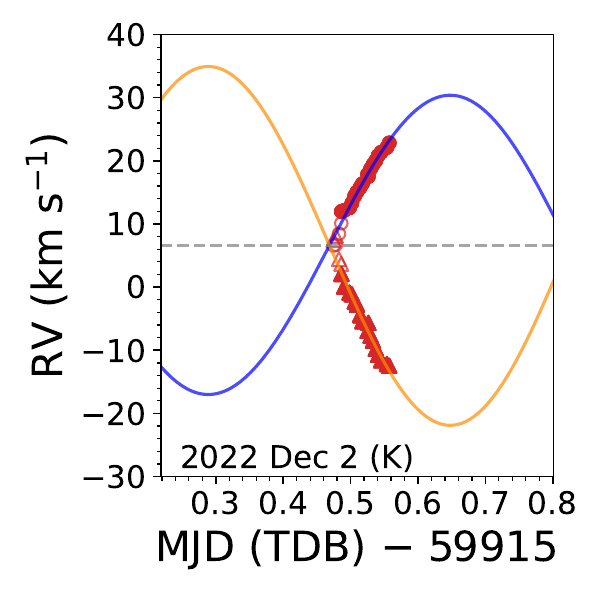} \\
\added{\includegraphics[width=0.32\textwidth]{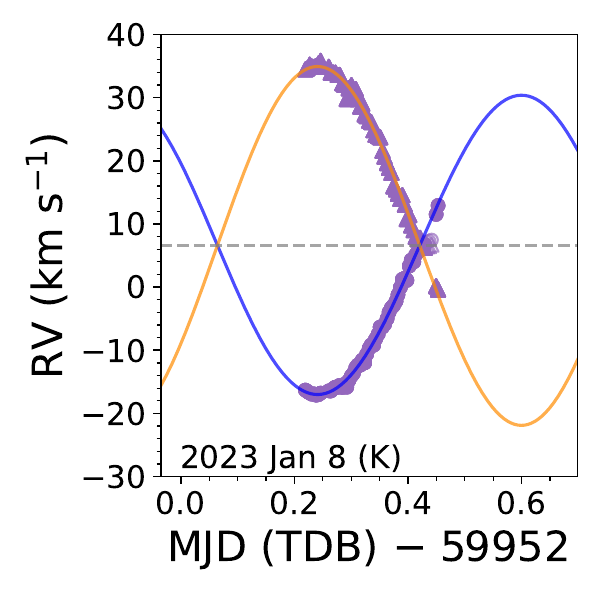}}
\includegraphics[width=0.64\textwidth]{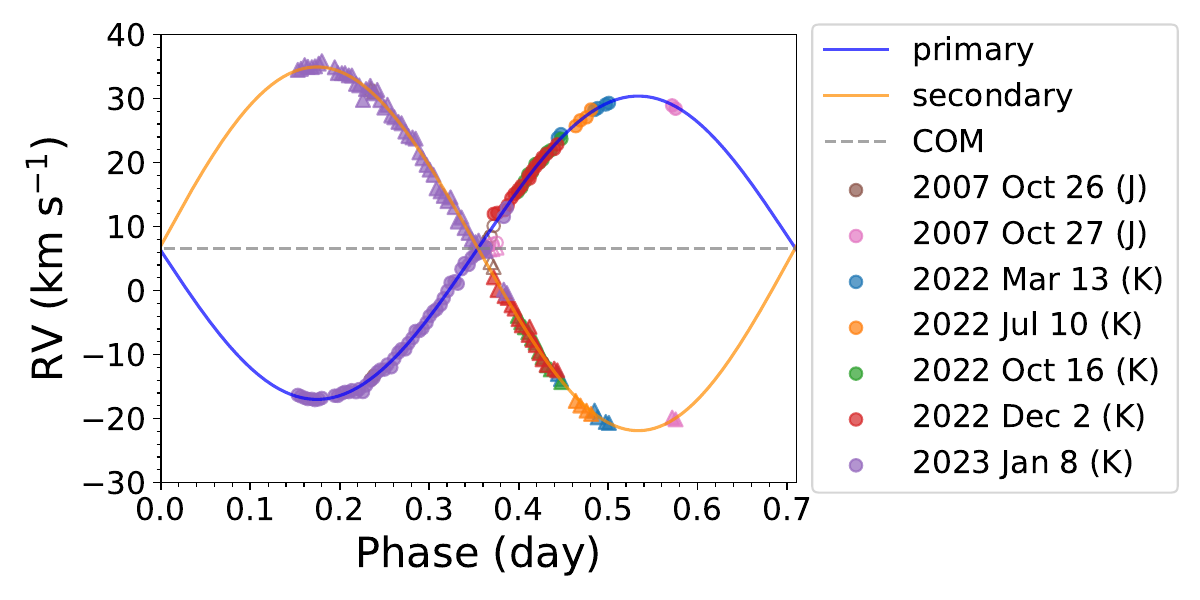} 
\vspace{-15pt}
\caption{\replaced{Radial velocity measurements on UT dates 2022 March 13, 2022 July 10, 2022 October 16, and 2022 December 2 (top and bottom-left panels), and the phase folded RV curve (bottom-right panel)
compared to the best-fit binary orbit.
RV measurements from $J$-band data on 2007 October 26 and 2007 October 27 are also plotted in the phase folded RV curve, but are not included in the orbit fit (see Section~\ref{sec:orbit}).
RV measurements for primary and secondary components are indicated by circles and triangles (uncertainties are smaller than the symbol size), while the corresponding 
orbit solutions are 
shown as blue and orange lines, respectively.
The center-of-mass (systematic) RV is indicated by the grey dashed line in all panels.}{Radial velocity measurements for primary (circles) and secondary (triangles) components for individual UT epochs 2007 October 26, 2007 October 27, and 2022 March 13 (top row); 2022 July 10, 2022 October 16, and 2022 December 2 (middle row); and 2023 January 8 (bottom left), and the phase folded RV curve (bottom right). Measurement uncertainties are smaller than the symbol size.
In each panel, the best-fit orbit indicating primary (blue lines) and secondary (orange lines) RVs are indicated, as well as the center-of-mass (systematic) RV (grey dashed lines).
The measurements from $J$-band data on 2007 October 26 and 2007 October 27 shown here have been corrected for the RV offset term $\sigma_J$.
Four epochs in 2022 December 2 and three epochs in 2023 January 8 were excluded from the orbit fit due to their unresolved spectra (open circles); these are nevertheless included in the phase-folded RV curve (see Section~\ref{sec:orbit}).}
\label{fig:orbit}}
\end{figure*}

The resulting best-fit orbit parameters \added{and their uncertainties} are listed in Table~\ref{tab:orbit}, and
the best-fit orbit with an overall reduced $\chi^2_r$ = \replaced{3.8}{2.5} is compared to the RV time series in Figure~\ref{fig:orbit}.
\replaced{Note that despite having four well-spaced observation sequences \added{in 2022}, all of the measurements span the same quarter phase of the orbit, which was verified through the asymmetric line depths.
This quarter phase is nevertheless well-sampled.}{Our observations sample over half of the orbit phase, and notably cover both oppositional points when the sources are at maximum RV separation.}
The best-fit period of the system of {\mper} is constrained to better than \replaced{1}{0.02}~s precision, \replaced{and}{although there may be unaccounted-for systematic errors due to our incomplete phase coverage. This period} 
is
\replaced{over three}{four} times shorter
than that of USCO~J1616$-$2512, making LP 413-53 the shortest-period UCD binary currently known.
We find a small but nonzero eccentricity, although our orbit sampling may bias the uncertainties for this parameter and we cannot rule out a fully circular orbit\added{ due to tidal circularization (see below)}.\deleted{Because of the low eccentricity, both $\omega$ and $M_0$ are poorly constrained.} 
The relative amplitudes of the primary and secondary RV curves allow us to directly measure the system mass ratio as $q = M_2/M_1 = K_1/K_2$ = \replaced{0.822$\pm$0.009}{\mq}, which is qualitatively consistent with the unequal line depths and inferred {\teff}s of the component spectra.
\added{We note that while our $\sigma_{J}$ is relatively large ({\sigmaj}), it is a sub-pixel shift, and we regard the $K$-band center-of-mass motion to be the reliable measure.}

\added{For completeness, we also determined orbital parameters by fitting the $K$-band data alone. We
found consistent orbital parameters of $P = 0.710623 \pm 0.000002$~day (note the 10 times larger uncertainty), $K_1 = 23.62 \pm 0.05$~{\kms}, $K_2 = 28.30 \pm 0.07$~{\kms}, $q = 0.8345 \pm 0.0017$, $v_0 = 6.61 \pm 0.02$~{\kms}, and $e = 0.0103 \pm 0.0017$; slightly different values of $M_0 = 1.52^{+0.18}_{-0.19}$~rad and $\omega = 1.65^{+0.19}_{-0.18}$~rad; and the same $\chi^2_r$ = 2.5.}

\deleted{Figure~\ref{fig:orbit} also displays the RVs inferred from the archival $J$-band data from 2007 October.
The NIRSPEC archival $J$-band RV measurements are highly inconsistent with the orbital solution from the $K$-band data, both in RV amplitude and orbital phases.
While the $J$-band data were better fit using a two-star solution compared to a single-star solution, the improvement is not as good, and there are larger and more structured residuals
in the spectral fits, indicating
imperfections in the theoretical stellar atmosphere models.
\added{We attempted to joint fit $J$- and $K$-band data and obtained similar parameters of $P = 0.710616 \pm 0.0000002$~day (0.6~s shorter), $K_1 = 23.79 \pm 0.05$~{\kms}, $K_2 = 28.53 \pm 0.06$~{\kms}, $q = 0.8336 \pm 0.0017$, $v_0 = 6.673 \pm 0.019$~{\kms}, a slightly higher $e = 0.0133 \pm 0.0018$, and different $M_0 = 2.01^{+0.09}_{-0.10}$~rad, and $\omega = 1.17^{+0.10}_{-0.09}$~rad, with reduced $\chi^2_r = 4.7$ (with $J$-band data) and reduced $\chi^2_r = 2.7$ (without $J$-band data).}
As the spectral model fits of the $K$-band CO bandhead $\sim$2.3~{\micron} are far more robust, 
\replaced{we restrict our analysis to these data and defer joint $J$- and $K$-band fits to a future study.}{we adopted the best-fit parameters using our $K$-band data for this study and provided these joint $J$- and $K$-band fits for reference.}}

\begin{deluxetable}{lc}
\tablecaption{System Parameters \label{tab:orbit}} 
\tabletypesize{\scriptsize} 
\tablehead{ 
\colhead{Parameter} &
\colhead{Value}
} 
\startdata 
\hline
\multicolumn{2}{c}{Spectral Fit Parameters} \\
\hline
$T_\mathrm{eff,1}$ &  \replaced{2884$\pm$42}{2879$\pm$37}~K \\
$T_\mathrm{eff,2}$ &  \replaced{2585$\pm$74}{2627$\pm$71}~K \\
{\logg}$_1$ &  \replaced{5.48$\pm$0.01}{5.49$\pm$0.02} (log cm/s$^2$)  \\
{\logg}$_2$  &  \replaced{5.46$\pm$0.04}{5.47$\pm$0.04} (log cm/s$^2$)  \\
{\vsini}$_1$ &  \replaced{5.7$\pm$1.2}{4.2$\pm$1.7}~{\kms} \\
{\vsini}$_2$ &  \replaced{6.1$\pm$1.8}{5.9$\pm$2.0}~{\kms} \\
$f_{2.3}$  &  \replaced{0.59$\pm$0.07}{0.65 $\pm$ 0.08} \\
\hline
\multicolumn{2}{c}{Orbit Fit Parameters} \\
\hline
$P$ & {\mper} \\
$K_1$  & {\mprv} \\
$K_2$  & {\msrv} \\
$\sqrt{e}\sin\omega$ & \esinw \\
$\sqrt{e}\cos\omega$ & \ecosw \\
$v_0$  & {\msysrv} \\
$M_0$  & {\manom} \\
$e$ & {\mecc} \\
$\omega$ & {\marg} \\
$q$ & {\mq} \\
$\sigma_{J}$ & {\sigmaj} \\
\hline
\multicolumn{2}{c}{Estimated Parameters} \\
\hline
Age & 1--10~Gyr \\
$T_\mathrm{eff,1}$ &  2395$\pm$133~K \\
$M_1$ & \masspri~$M_\odot$ \\
$M_2$ & \masssec~$M_\odot$ \\
$R_{1,2}$ & \radii~$R_\odot$ \\
$a$  & \separation~au \\
  & \sepradii~R$_{1,2}$ \\
$i$  & \inc$\degr$ \\
\enddata
\end{deluxetable}

\section{Discussion} \label{sec:discuss}


The short orbital period of LP 413-53 makes it a likely candidate for eclipsing, depending on its physical separation and component radii, with the latter depending on the system age.
We re-evaluated the kinematic evidence of Hyades association membership by combining
our systematic RV
with \textit{Gaia} DR3 astrometry, 
which yields a Galactic heliocentric velocity of 
($U$, $V$, $W$) = \replaced{($-17.64 \pm 0.09$, $-24.66 \pm 0.11$, $+12.37 \pm 0.08$ )}{($-17.52 \pm 0.05$, $-24.68 \pm 0.11$, $+12.43 \pm 0.07$)}~{\kms}.
The velocity and position of the LP 413-53 system exclude\deleted{s} membership in the Hyades, and in 26 other young, nearby clusters and associations, based on the
BANYAN $\Sigma$ webtool\footnote{\url{https://www.exoplanetes.umontreal.ca/banyan/banyansigma.php}.} \citep{Gagne:2018ab}.
The (occasionally) strong magnetic activity of LP~413-53 does not provide a firm constraint on the age of the system given the long magnetic activity lifetimes of very-low-mass stars \citep{West:2008aa}, particularly binaries \citep{2012AJ....144...93M}.
We therefore assume for the rest of this analysis that the components of LP~413-53 are low-mass stars with ages $\gtrsim$1~Gyr.

Component masses and radii were estimated from evolutionary models assuming that the combined-light M9 optical classification of the system reflects the primary component. We used the spectral type to {\teff} relation from \cite{Filippazzo:2015aa} to estimate a primary {\teff} = 2395$\pm$133~K, which is consistent to fits of broadband photometry to stellar models (2334$^{+32}_{-24}$~K using stellar models from \citealt{2013A&A...553A...6H}). For an age of 1--10~Gyr, this temperature corresponds to a mass of 0.076--0.087~M$_\odot$ and radius\footnote{We are not taking into account 10\%--20\% radius inflation that is well-documented among very-low-mass stars \citep{2018AJ....155..225K}.} of 0.091--0.101~R$_\odot$ 
using the \cite{Burrows:2001aa} models and 
0.076--\replaced{0.087}{0.086}~M$_\odot$ and 0.098-0.109~R$_\odot$ using the \cite{Baraffe:2003aa} models. 
Choosing the outer range\added{s} \replaced{of}{encompassing} these values, our inferred mass ratio implies a secondary mass of \masssec~M$_\odot$, making it a possible brown dwarf.\footnote{The lower range of secondary mass is consistent with the absence of Li~I absorption in \added{combined-light} optical spectra \citep{Reid:1999aa}.}
These values yield a total system mass of 0.14--0.16~M$_\odot$,
which combined with the period implies a semi-major axis $a$ = \separation~au
or {\sepradii} stellar radii\deleted{ encompassing the range of estimates}.
We can further estimate the system inclination from the expected primary RV amplitude given these parameters:
\begin{equation}
    K_1 = \frac{2\pi{a}}{P}\frac{q}{1+q}\frac{\sin{i}}{\sqrt{1-e^2}} \approx \replaced{54.0}{57.5}~{\rm km/s}~\sin{i}
\end{equation}
Comparing this value
to our observed \deleted{value} $K_1$ = \replaced{24.2}{23.7}~{\kms} yields an estimated inclination angle of \replaced{27}{\incshort}$\degr$ and a corresponding projected separation on the sky of \replaced{17--20}{7--8} stellar radii at closest approach. Thus, it is unlikely that this system will eclipse assuming the estimated component parameters are accurate, although observational confirmation of this assessment is warranted. 
We note that LP 413-53 is too faint for \textit{TESS} monitoring and was not targeted with \textit{Kepler/K2}.

While the estimated physical properties of the components of this system indicate that they do not currently come into close physical contact, this may not have been the case if the system formed at its current separation.
The pre-Main Sequence (ages $\sim$ 1~Myr) radii of very-low-mass stars are several times larger than their evolved configurations \citep{Burrows:2001aa, Baraffe:2003aa, 2006Natur.440..311S}, bringing the components to near-contact at early ages. 
Like more massive short-period stellar binaries, this system likely underwent orbital decay from an initially wider separation via angular momentum loss through tidal friction and magnetized winds \citep{Stepien:1995aa}. 
Alternately, LP~413-53 could have formed as part of an unstable triple or higher order system, with the third component carrying away 
angular momentum from the now hardened pair.
Indeed, several close UCD pairs, including SPEC~1510$-$2828AB \citep{Triaud:2020aa} and SDSS J0006$-$0852AB \citep{Burgasser:2012aa} are found in (presumed stable) hierarchical triples.
In the case of LP~413-53, the absence of a co-moving companion within 1$\degr$ (0.6~pc) would require this component to be fully ejected from the system.
Both cases are expected to produce a low eccentricity, synchronously rotating binary given the short tidal circularization \replaced{($\sim4\times10^5$~yr)}{($\sim2\times10^5$~yr)}
and synchronization  \replaced{($\sim6\times10^3$~yr)}{($\sim3\times10^3$~yr)} 
timescales for fully convective stars at this separation \citep{Zahn:1977aa,2019ApJ...881...88F}.
The corresponding synchronized equatorial rotational velocity, $\sim$\replaced{6}{7}~{\kms} for a fully evolved low-mass star\added{, consistent with projected {\vsini} $\sim$ 3~{\kms} if the spin axis is aligned with the orbital plane axis}, is consistent with the observed values \replaced{that are close to}{given our} detection limits.
Higher resolution data could provide a second constraint on the orbit geometry assuming spin-orbit alignment, or alternately test that alignment and potentially differentiate between formation via slow tidal dissipation or three-body interaction.

Finally, we note that while no physical interaction is present in this system, magnetospheric interaction, even indirect, likely is. 
The radio magnetospheres of low-mass stars extend 1--3 stellar radii beyond the photosphere \citep{1997A&A...317..707A,2011ApJ...735L...2R,2013ApJ...762L...3B},
and can be further distorted by the gravitational potential of the system. 
Such interaction may explain the highly variable H$\alpha$ measurements reported in the literature, and warrants optical spectral monitoring to search for phased emission as seen in other short-period M dwarf systems (cf.\ \citealt{2013ApJ...764...62D}).
We note that the long convective timescales of the lowest-mass stars ($\gtrsim$100~day; \citealt{2011ApJ...743...48W}) \replaced{implies}{imply} that both components will remain in the saturated regime for magnetic emission (Ro $<$ 10$^{-2}$; \citealt{2017ApJ...834...85N}), and are thus likely to continue to lose angular momentum through dual magnetized winds.

\begin{acknowledgments}

The authors thank observing assistants Heather Hershley, Anthony Connors, Julie Renaud-Kim, and Matthew Wahl, and supporting astronomers \replaced{Jim}{James} Lyke, Rosalie McGurk, Percy Gomez, and Gregory Doppmann for their help in obtaining our Keck/NIRSPEC observations on 2022 March 13 (UT), 2022 July 10 (UT), \replaced{and 2022 October 16 (UT)}{2022 October 16 (UT), 2022 December 2 (UT), and 2023 January 8 (UT)}.
The authors \deleted{would} also like to thank Erik Petigura and Dakotah Tyler for their generous swap in the Keck/NIRSPEC observation on 2022 October 16\deleted{, during which we obtained robust RV which provided better constraint on the RV binary orbital phase}.
\added{The authors thank the anonymous referee for his/her/their valuable review which has improved the original manuscript.}
Data presented herein were obtained at the W. M. Keck Observatory, which is operated as a scientific partnership among the California Institute of Technology, the University of California, and the National Aeronautics and Space Administration. The Observatory was made possible by the generous financial support of the W. M. Keck Foundation. 
This research has made use of the Keck Observatory Archive (KOA), which is operated by the W. M. Keck Observatory and the NASA Exoplanet Science Institute (NExScI), under contract with the National Aeronautics and Space Administration.
%
The authors recognize and acknowledge the significant cultural role
and reverence that the summit of Maunakea has with the
indigenous Hawaiian community, and that the W. M. Keck
Observatory stands on Crown and Government Lands that the
State of Hawai'i is obligated to protect and preserve for future
generations of indigenous Hawaiians. 
%
%
Portions of this work
were conducted at the University of California, San Diego,
which was built on the unceded territory of the Kumeyaay
Nation, whose people continue to maintain their political
sovereignty and cultural traditions as vital members of the San
Diego community.

\end{acknowledgments}

\vspace{5mm}
\facilities{Keck:II (NIRSPEC)}

\software{\texttt{Astropy} \citep{Astropy-Collaboration:2018aa}, 
\texttt{BANYAN $\Sigma$} \citep{Gagne:2018ab},
\texttt{SMART} \citep{Hsu:2021aa},
\texttt{SPLAT} \citep{Burgasser:2017ac}
          }

\begin{longrotatetable}
\begin{deluxetable}{llccccccccccccc}
\tablecaption{NIRSPEC Measurements \label{tab:nirspec}} 
\tabletypesize{\scriptsize} 
\tablehead{ 
 & & & & \multicolumn{4}{c}{Primary} & \multicolumn{4}{c}{Secondary} & \\
\cline{5-8} \cline{9-12}
 &  &  &  & 
\colhead{{\teff}} & 
\colhead{{\logg}} & 
\colhead{RV}  & 
\colhead{{\vsini}}  &
\colhead{{\teff}} & 
\colhead{{\logg}} & 
\colhead{RV}  & 
\colhead{{\vsini}}  &
\colhead{{C$_\mathrm{scale}$\tablenotemark{b}}}  &
\colhead{$\chi^2_r$\tablenotemark{c}} &
\colhead{$\chi^2_{r, s}$\tablenotemark{c, d}} \\ 
\colhead{UT Date} &
\colhead{MJD\added{ (TDB)}\tablenotemark{a}} &
\colhead{S/N} &
\colhead{order} &
\colhead{(K)} & 
\colhead{(cm s$^{-2}$)} & 
\colhead{(km s$^{-1}$)}  & 
\colhead{(km s$^{-1}$)}  & 
\colhead{(K)} & 
\colhead{(cm s$^{-2}$)} & 
\colhead{(km s$^{-1}$)}  & 
\colhead{(km s$^{-1}$)}  &  \\ 
} 
\startdata 
\replaced{2007 Oct 26 & 54399.54477 & 18.8 & 57 & $3027^{+40}_{-42}$ & $4.74^{+0.26}_{-0.38}$ & $-11.43^{+0.25}_{-0.23}$ & $4.3^{+1.14}_{-1.33}$ & $2858^{+50}_{-100}$ & $4.74^{+0.26}_{-0.39}$ & $33.94^{+0.58}_{-0.4}$ & $5.14^{+1.88}_{-2.22}$ & $0.55^{+0.05}_{-0.05}$ & 2.6 & 4.4 \\ 
2007 Oct 26 & 54399.54852 & 19.1 & 57 & $3001^{+33}_{-32}$ & $4.88^{+0.2}_{-0.2}$ & $-10.8^{+0.29}_{-0.22}$ & $4.11^{+1.02}_{-2.25}$ & $2962^{+50}_{-62}$ & $4.88^{+0.42}_{-0.44}$ & $33.57^{+0.35}_{-0.34}$ & $3.1^{+1.78}_{-1.57}$ & $0.56^{+0.05}_{-0.04}$ & 2.6 & 4.1 \\ 
2007 Oct 27 & 54400.59211 & 16.5 & 57 & $2944^{+37}_{-38}$ & $5.42^{+0.05}_{-0.08}$ & $31.4^{+0.35}_{-0.25}$ & $4.43^{+1.16}_{-1.18}$ & $2978^{+76}_{-72}$ & $5.42^{+0.24}_{-0.23}$ & $-17.36^{+0.48}_{-0.55}$ & $6.93^{+1.07}_{-1.6}$ & $0.68^{+0.05}_{-0.06}$ & 2.4 & 3.8 \\ 
2007 Oct 27 & 54400.59595 & 16.5 & 57 & $2920^{+37}_{-32}$ & $4.91^{+0.2}_{-0.18}$ & $30.89^{+0.29}_{-0.27}$ & $5.21^{+1.67}_{-1.42}$ & $2924^{+60}_{-74}$ & $4.91^{+0.43}_{-0.44}$ & $-17.64^{+0.41}_{-0.46}$ & $3.57^{+1.59}_{-1.65}$ & $0.64^{+0.06}_{-0.05}$ & 2.5 & 4.1 \\ 
2022 Mar 13 & 59651.21041 & 22.3 & 33 & $2895^{+15}_{-11}$ & $5.49^{+0.01}_{-0.01}$ & $23.84^{+0.13}_{-0.16}$ & $6.55^{+0.42}_{-0.52}$ & $2617^{+36}_{-35}$ & $5.49^{+0.02}_{-0.03}$ & $-13.04^{+0.28}_{-0.27}$ & $4.56^{+0.9}_{-1.28}$ & $0.57^{+0.01}_{-0.01}$ & 2.7 & 7.6 \\ 
2022 Mar 13 & 59651.21388 & 24.2 & 33 & $2861^{+14}_{-12}$ & $5.49^{+0.0}_{-0.01}$ & $24.43^{+0.14}_{-0.14}$ & $6.72^{+0.32}_{-0.62}$ & $2639^{+36}_{-31}$ & $5.49^{+0.01}_{-0.02}$ & $-13.83^{+0.16}_{-0.17}$ & $3.31^{+1.21}_{-1.86}$ & $0.57^{+0.01}_{-0.01}$ & 3.0 & 9.1 \\ 
2022 Mar 13 & 59651.25139 & 29.2 & 33 & $2893^{+12}_{-12}$ & $5.49^{+0.01}_{-0.01}$ & $28.22^{+0.12}_{-0.11}$ & $5.49^{+0.71}_{-0.7}$ & $2604^{+35}_{-38}$ & $5.49^{+0.03}_{-0.07}$ & $-18.75^{+0.24}_{-0.23}$ & $6.61^{+0.86}_{-0.73}$ & $0.6^{+0.01}_{-0.01}$ & 3.6 & 10.9 \\ 
2022 Mar 13 & 59651.25486 & 29.9 & 33 & $2874^{+11}_{-12}$ & $5.5^{+0.0}_{-0.01}$ & $28.5^{+0.16}_{-0.12}$ & $6.41^{+0.21}_{-0.64}$ & $2631^{+30}_{-33}$ & $5.5^{+0.01}_{-0.01}$ & $-19.82^{+0.2}_{-0.22}$ & $6.95^{+0.9}_{-0.93}$ & $0.6^{+0.02}_{-0.01}$ & 4.0 & 11.6 \\ 
2022 Mar 13 & 59651.26395 & 22.8 & 33 & $2905^{+18}_{-17}$ & $5.49^{+0.02}_{-0.01}$ & $29.07^{+0.14}_{-0.2}$ & $6.66^{+0.33}_{-0.48}$ & $2645^{+40}_{-37}$ & $5.49^{+0.03}_{-0.04}$ & $-20.52^{+0.3}_{-0.27}$ & $4.59^{+0.97}_{-2.2}$ & $0.58^{+0.02}_{-0.02}$ & 2.8 & 7.5 \\ 
2022 Mar 13 & 59651.26742 & 22.5 & 33 & $2883^{+13}_{-13}$ & $5.49^{+0.01}_{-0.01}$ & $29.31^{+0.16}_{-0.18}$ & $5.26^{+0.5}_{-0.52}$ & $2615^{+28}_{-33}$ & $5.49^{+0.02}_{-0.03}$ & $-20.68^{+0.22}_{-0.25}$ & $6.3^{+1.2}_{-1.31}$ & $0.61^{+0.02}_{-0.02}$ & 2.9 & 7.4 \\ 
2022 Jul 10 & 59770.61545 & 9.7 & 33 & $2892^{+31}_{-28}$ & $5.48^{+0.02}_{-0.04}$ & $25.69^{+0.27}_{-0.24}$ & $8.29^{+0.87}_{-0.95}$ & $2519^{+88}_{-79}$ & $5.24^{+0.19}_{-0.19}$ & $-17.25^{+0.49}_{-0.3}$ & $8.0^{+2.36}_{-2.16}$ & $0.6^{+0.05}_{-0.03}$ & 1.8 & 3.0 \\ 
2022 Jul 10 & 59770.62066 & 14.8 & 33 & $2888^{+22}_{-22}$ & $5.48^{+0.01}_{-0.02}$ & $26.59^{+0.23}_{-0.22}$ & $7.47^{+0.61}_{-0.74}$ & $2639^{+34}_{-48}$ & $5.46^{+0.03}_{-0.05}$ & $-17.96^{+0.34}_{-0.31}$ & $4.63^{+1.71}_{-1.57}$ & $0.6^{+0.03}_{-0.02}$ & 2.2 & 4.6 \\ 
2022 Jul 10 & 59770.6275 & 14.8 & 33 & $2884^{+26}_{-17}$ & $5.48^{+0.01}_{-0.03}$ & $27.06^{+0.16}_{-0.16}$ & $7.2^{+0.69}_{-0.6}$ & $2579^{+50}_{-80}$ & $5.44^{+0.05}_{-0.05}$ & $-18.76^{+0.28}_{-0.37}$ & $6.39^{+1.44}_{-1.6}$ & $0.63^{+0.02}_{-0.02}$ & 2.2 & 4.4 \\ 
2022 Jul 10 & 59770.63271 & 17.9 & 33 & $2880^{+21}_{-19}$ & $5.48^{+0.01}_{-0.03}$ & $28.28^{+0.24}_{-0.19}$ & $7.05^{+0.81}_{-0.92}$ & $2600^{+39}_{-41}$ & $5.45^{+0.03}_{-0.08}$ & $-19.28^{+0.22}_{-0.24}$ & $4.53^{+1.69}_{-1.67}$ & $0.58^{+0.02}_{-0.02}$ & 2.6 & 4.0 \\ 
2022 Oct 16 & 59868.60714 & 22.3 & 33 & $2872^{+16}_{-15}$ & $5.49^{+0.01}_{-0.01}$ & $15.41^{+0.16}_{-0.11}$ & $5.09^{+0.39}_{-0.46}$ & $2642^{+40}_{-42}$ & $5.48^{+0.02}_{-0.03}$ & $-3.97^{+0.21}_{-0.24}$ & $2.86^{+1.19}_{-1.54}$ & $0.57^{+0.03}_{-0.02}$ & 3.4 & 5.9 \\ 
2022 Oct 16 & 59868.61104 & 18.2 & 33 & $2867^{+18}_{-20}$ & $5.49^{+0.01}_{-0.02}$ & $16.07^{+0.14}_{-0.14}$ & $5.17^{+0.44}_{-0.53}$ & $2676^{+36}_{-38}$ & $5.47^{+0.02}_{-0.03}$ & $-4.96^{+0.31}_{-0.27}$ & $6.48^{+0.56}_{-0.98}$ & $0.58^{+0.03}_{-0.02}$ & 2.7 & 5.0 \\ 
2022 Oct 16 & 59868.61511 & 23.8 & 33 & $2885^{+19}_{-20}$ & $5.49^{+0.01}_{-0.02}$ & $17.07^{+0.16}_{-0.13}$ & $5.19^{+0.29}_{-0.32}$ & $2638^{+49}_{-32}$ & $5.48^{+0.01}_{-0.03}$ & $-5.74^{+0.23}_{-0.25}$ & $3.9^{+1.46}_{-1.41}$ & $0.6^{+0.03}_{-0.02}$ & 3.4 & 6.8 \\ 
2022 Oct 16 & 59868.619 & 22.7 & 33 & $2910^{+16}_{-19}$ & $5.49^{+0.01}_{-0.01}$ & $18.16^{+0.14}_{-0.13}$ & $5.33^{+0.48}_{-0.65}$ & $2552^{+51}_{-37}$ & $5.48^{+0.01}_{-0.01}$ & $-6.52^{+0.31}_{-0.25}$ & $4.55^{+0.86}_{-1.21}$ & $0.61^{+0.02}_{-0.02}$ & 3.4 & 8.6 \\ 
2022 Oct 16 & 59868.62304 & 21.1 & 33 & $2914^{+18}_{-11}$ & $5.49^{+0.01}_{-0.01}$ & $18.47^{+0.15}_{-0.13}$ & $4.48^{+0.74}_{-0.95}$ & $2588^{+36}_{-30}$ & $5.48^{+0.01}_{-0.03}$ & $-7.48^{+0.2}_{-0.25}$ & $6.31^{+1.11}_{-0.74}$ & $0.65^{+0.02}_{-0.03}$ & 3.1 & 6.2 \\ 
2022 Oct 16 & 59868.62693 & 25.4 & 33 & $2977^{+15}_{-15}$ & $5.49^{+0.01}_{-0.01}$ & $19.75^{+0.12}_{-0.16}$ & $5.56^{+0.77}_{-0.76}$ & $2358^{+7}_{-18}$ & $5.47^{+0.02}_{-0.04}$ & $-8.13^{+0.24}_{-0.26}$ & $3.45^{+1.08}_{-1.41}$ & $0.7^{+0.04}_{-0.02}$ & 3.6 & 10.2 \\ 
2022 Oct 16 & 59868.63099 & 20.4 & 33 & $2886^{+18}_{-17}$ & $5.49^{+0.01}_{-0.02}$ & $19.96^{+0.21}_{-0.12}$ & $5.77^{+0.76}_{-0.85}$ & $2603^{+50}_{-41}$ & $5.47^{+0.02}_{-0.03}$ & $-9.51^{+0.19}_{-0.23}$ & $4.29^{+1.27}_{-1.45}$ & $0.62^{+0.02}_{-0.02}$ & 2.9 & 6.1 \\ 
2022 Oct 16 & 59868.63485 & 22.7 & 33 & $2901^{+19}_{-20}$ & $5.49^{+0.01}_{-0.01}$ & $20.45^{+0.14}_{-0.1}$ & $5.11^{+0.47}_{-0.33}$ & $2636^{+29}_{-46}$ & $5.49^{+0.01}_{-0.02}$ & $-10.45^{+0.18}_{-0.22}$ & $4.53^{+1.29}_{-1.03}$ & $0.6^{+0.02}_{-0.01}$ & 3.2 & 7.7 \\ 
2022 Oct 16 & 59868.64011 & 14.4 & 33 & $2902^{+25}_{-38}$ & $5.47^{+0.02}_{-0.05}$ & $21.57^{+0.31}_{-0.23}$ & $6.08^{+0.62}_{-0.99}$ & $2560^{+72}_{-62}$ & $5.43^{+0.05}_{-0.07}$ & $-11.7^{+0.36}_{-0.35}$ & $6.68^{+1.51}_{-1.39}$ & $0.66^{+0.02}_{-0.03}$ & 2.2 & 4.2 \\ 
2022 Oct 16 & 59868.64397 & 20.1 & 33 & $2919^{+18}_{-17}$ & $5.49^{+0.01}_{-0.02}$ & $21.91^{+0.17}_{-0.13}$ & $5.37^{+0.27}_{-0.53}$ & $2616^{+48}_{-42}$ & $5.48^{+0.01}_{-0.02}$ & $-12.33^{+0.32}_{-0.25}$ & $3.52^{+1.28}_{-1.54}$ & $0.57^{+0.02}_{-0.02}$ & 2.8 & 6.7 \\ 
2022 Oct 16 & 59868.65584 & 7.7 & 33 & $2955^{+29}_{-42}$ & $5.46^{+0.03}_{-0.06}$ & $23.7^{+0.26}_{-0.17}$ & $3.29^{+1.03}_{-1.4}$ & $2595^{+54}_{-71}$ & $5.43^{+0.05}_{-0.1}$ & $-14.31^{+0.41}_{-0.4}$ & $5.86^{+1.96}_{-1.8}$ & $0.66^{+0.04}_{-0.04}$ & 1.5 & 2.4 \\ 
2022 Dec 2 & 59915.46833 & 15.0 & 33 & $2930^{+28}_{-30}$ & $5.49^{+0.01}_{-0.02}$ & $6.58^{+0.27}_{-0.26}$ & $2.1^{+1.03}_{-1.27}$ & $2561^{+58}_{-94}$ & $5.49^{+0.01}_{-0.02}$ & $7.47^{+0.58}_{-0.53}$ & $9.36^{+0.46}_{-0.85}$ & $0.85^{+0.1}_{-0.17}$ & 2.4 & 3.2 \\ 
2022 Dec 2 & 59915.4718 & 12.7 & 33 & $2893^{+31}_{-36}$ & $5.49^{+0.01}_{-0.02}$ & $6.7^{+0.57}_{-0.68}$ & $5.95^{+0.83}_{-1.08}$ & $2621^{+66}_{-53}$ & $5.48^{+0.01}_{-0.03}$ & $8.13^{+1.14}_{-0.97}$ & $8.67^{+1.06}_{-2.11}$ & $0.8^{+0.14}_{-0.2}$ & 2.0 & 2.4 \\ 
2022 Dec 2 & 59915.4764 & 10.5 & 33 & $2911^{+50}_{-56}$ & $5.48^{+0.02}_{-0.03}$ & $8.44^{+0.7}_{-0.78}$ & $6.78^{+1.06}_{-0.84}$ & $2584^{+121}_{-113}$ & $5.47^{+0.02}_{-0.04}$ & $4.44^{+1.53}_{-1.2}$ & $8.65^{+1.1}_{-1.58}$ & $0.75^{+0.16}_{-0.16}$ & 2.0 & 2.4 \\ 
2022 Dec 2 & 59915.47987 & 9.2 & 33 & $2820^{+51}_{-41}$ & $5.48^{+0.01}_{-0.03}$ & $10.11^{+0.62}_{-0.71}$ & $7.67^{+1.29}_{-1.42}$ & $2737^{+45}_{-68}$ & $5.47^{+0.02}_{-0.05}$ & $3.69^{+1.04}_{-1.08}$ & $8.54^{+1.09}_{-1.84}$ & $0.75^{+0.16}_{-0.17}$ & 1.6 & 2.0 \\ 
2022 Dec 2 & 59915.48 & 15.7 & 33 & $2886^{+36}_{-34}$ & $5.48^{+0.01}_{-0.04}$ & $11.97^{+0.39}_{-0.35}$ & $3.94^{+1.21}_{-0.66}$ & $2649^{+50}_{-45}$ & $5.47^{+0.02}_{-0.03}$ & $2.08^{+0.63}_{-0.58}$ & $6.03^{+1.57}_{-1.34}$ & $0.77^{+0.1}_{-0.16}$ & 2.0 & 2.5 \\ 
2022 Dec 2 & 59915.48411 & 14.2 & 33 & $2755^{+39}_{-31}$ & $5.48^{+0.02}_{-0.02}$ & $12.14^{+0.47}_{-0.35}$ & $6.26^{+1.09}_{-0.95}$ & $2818^{+62}_{-52}$ & $5.47^{+0.02}_{-0.04}$ & $0.08^{+0.62}_{-0.51}$ & $6.06^{+1.33}_{-1.38}$ & $0.59^{+0.13}_{-0.07}$ & 2.1 & 2.6 \\ 
2022 Dec 2 & 59915.49204 & 13.3 & 33 & $2838^{+22}_{-22}$ & $5.48^{+0.01}_{-0.02}$ & $12.56^{+0.24}_{-0.24}$ & $4.34^{+0.88}_{-0.99}$ & $2671^{+57}_{-51}$ & $5.45^{+0.04}_{-0.06}$ & $-0.87^{+0.43}_{-0.39}$ & $4.44^{+1.7}_{-1.29}$ & $0.54^{+0.06}_{-0.03}$ & 2.1 & 2.6 \\ 
2022 Dec 2 & 59915.49551 & 11.7 & 33 & $2865^{+32}_{-31}$ & $5.49^{+0.01}_{-0.01}$ & $13.31^{+0.24}_{-0.22}$ & $5.24^{+0.52}_{-0.73}$ & $2619^{+63}_{-75}$ & $5.48^{+0.02}_{-0.03}$ & $-1.14^{+0.43}_{-0.38}$ & $3.83^{+1.51}_{-2.11}$ & $0.58^{+0.08}_{-0.04}$ & 1.8 & 2.3 \\ 
2022 Dec 2 & 59915.49975 & 10.8 & 33 & $2910^{+25}_{-35}$ & $5.48^{+0.02}_{-0.04}$ & $14.46^{+0.22}_{-0.26}$ & $4.86^{+0.73}_{-0.69}$ & $2605^{+67}_{-59}$ & $5.45^{+0.03}_{-0.05}$ & $-2.3^{+0.36}_{-0.34}$ & $2.68^{+1.44}_{-1.91}$ & $0.57^{+0.04}_{-0.04}$ & 1.9 & 2.4 \\ 
2022 Dec 2 & 59915.50322 & 9.2 & 33 & $2833^{+29}_{-35}$ & $5.48^{+0.02}_{-0.03}$ & $15.09^{+0.3}_{-0.31}$ & $6.74^{+0.89}_{-0.81}$ & $2689^{+66}_{-58}$ & $5.46^{+0.03}_{-0.08}$ & $-2.82^{+0.37}_{-0.4}$ & $4.47^{+1.74}_{-1.24}$ & $0.59^{+0.05}_{-0.04}$ & 1.8 & 2.3 \\ 
2022 Dec 2 & 59915.50772 & 17.5 & 33 & $2826^{+22}_{-23}$ & $5.49^{+0.01}_{-0.01}$ & $15.77^{+0.22}_{-0.21}$ & $6.05^{+0.59}_{-0.42}$ & $2724^{+35}_{-26}$ & $5.48^{+0.01}_{-0.03}$ & $-4.45^{+0.28}_{-0.2}$ & $5.19^{+0.92}_{-1.21}$ & $0.6^{+0.02}_{-0.04}$ & 2.6 & 3.8 \\ 
2022 Dec 2 & 59915.51119 & 19.2 & 33 & $2872^{+18}_{-20}$ & $5.49^{+0.01}_{-0.02}$ & $16.38^{+0.14}_{-0.11}$ & $6.12^{+0.47}_{-0.83}$ & $2660^{+46}_{-52}$ & $5.48^{+0.02}_{-0.03}$ & $-5.43^{+0.26}_{-0.26}$ & $6.23^{+0.66}_{-0.8}$ & $0.58^{+0.02}_{-0.02}$ & 2.6 & 4.4 \\ 
2022 Dec 2 & 59915.52 & 19.8 & 33 & $2875^{+16}_{-20}$ & $5.47^{+0.02}_{-0.03}$ & $17.5^{+0.15}_{-0.16}$ & $4.93^{+0.58}_{-0.28}$ & $2637^{+43}_{-51}$ & $5.47^{+0.02}_{-0.04}$ & $-5.67^{+0.28}_{-0.25}$ & $2.48^{+1.58}_{-1.59}$ & $0.55^{+0.02}_{-0.02}$ & 2.7 & 4.9 \\ 
2022 Dec 2 & 59915.51892 & 17.3 & 33 & $2824^{+24}_{-24}$ & $5.49^{+0.01}_{-0.02}$ & $17.89^{+0.17}_{-0.14}$ & $5.53^{+0.76}_{-0.78}$ & $2745^{+31}_{-42}$ & $5.48^{+0.02}_{-0.04}$ & $-6.92^{+0.18}_{-0.27}$ & $6.3^{+0.84}_{-0.79}$ & $0.63^{+0.02}_{-0.03}$ & 2.7 & 4.6 \\ 
2022 Dec 2 & 59915.52336 & 17.7 & 33 & $2854^{+21}_{-18}$ & $5.48^{+0.01}_{-0.02}$ & $18.74^{+0.14}_{-0.21}$ & $6.17^{+0.42}_{-0.55}$ & $2722^{+39}_{-42}$ & $5.48^{+0.02}_{-0.04}$ & $-7.71^{+0.32}_{-0.29}$ & $5.92^{+0.87}_{-1.21}$ & $0.59^{+0.02}_{-0.02}$ & 2.8 & 4.9 \\ 
2022 Dec 2 & 59915.52683 & 16.6 & 33 & $2876^{+19}_{-18}$ & $5.49^{+0.01}_{-0.02}$ & $19.39^{+0.13}_{-0.11}$ & $6.32^{+0.39}_{-0.53}$ & $2627^{+49}_{-41}$ & $5.47^{+0.02}_{-0.04}$ & $-8.5^{+0.3}_{-0.25}$ & $6.37^{+1.06}_{-0.85}$ & $0.59^{+0.02}_{-0.02}$ & 2.3 & 4.5 \\ 
2022 Dec 2 & 59915.53107 & 15.8 & 33 & $2858^{+23}_{-29}$ & $5.49^{+0.01}_{-0.02}$ & $20.03^{+0.19}_{-0.13}$ & $5.48^{+0.88}_{-0.66}$ & $2693^{+44}_{-58}$ & $5.48^{+0.02}_{-0.04}$ & $-9.76^{+0.26}_{-0.27}$ & $4.19^{+1.16}_{-1.82}$ & $0.57^{+0.02}_{-0.02}$ & 2.1 & 4.4 \\ 
2022 Dec 2 & 59915.53454 & 14.5 & 33 & $2807^{+29}_{-24}$ & $5.49^{+0.01}_{-0.02}$ & $20.8^{+0.18}_{-0.18}$ & $6.66^{+0.6}_{-0.6}$ & $2705^{+53}_{-49}$ & $5.44^{+0.04}_{-0.05}$ & $-10.74^{+0.34}_{-0.33}$ & $7.95^{+1.0}_{-0.79}$ & $0.62^{+0.03}_{-0.02}$ & 2.5 & 5.0 \\ 
2022 Dec 2 & 59915.53901 & 13.3 & 33 & $2799^{+26}_{-29}$ & $5.48^{+0.01}_{-0.02}$ & $21.36^{+0.19}_{-0.18}$ & $4.43^{+0.88}_{-1.04}$ & $2750^{+51}_{-42}$ & $5.46^{+0.03}_{-0.04}$ & $-11.6^{+0.27}_{-0.3}$ & $8.31^{+0.83}_{-0.77}$ & $0.67^{+0.03}_{-0.03}$ & 2.3 & 4.3 \\ 
2022 Dec 2 & 59915.5475 & 13.6 & 33 & $2885^{+26}_{-20}$ & $5.49^{+0.01}_{-0.01}$ & $22.12^{+0.14}_{-0.18}$ & $5.51^{+0.85}_{-0.74}$ & $2608^{+51}_{-61}$ & $5.45^{+0.04}_{-0.04}$ & $-12.2^{+0.38}_{-0.32}$ & $4.54^{+1.37}_{-1.39}$ & $0.55^{+0.02}_{-0.02}$ & 2.0 & 3.9 \\ 
2022 Dec 2 & 59915.55097 & 15.0 & 33 & $2926^{+24}_{-18}$ & $5.48^{+0.01}_{-0.02}$ & $22.83^{+0.18}_{-0.22}$ & $5.39^{+0.42}_{-0.59}$ & $2643^{+43}_{-50}$ & $5.47^{+0.02}_{-0.04}$ & $-12.46^{+0.29}_{-0.29}$ & $7.35^{+1.06}_{-1.06}$ & $0.58^{+0.02}_{-0.02}$ & 2.3 & 5.1 \\}
{2007 Oct 26\tablenotemark{e} & 54399.55066 & 18.8 & 57 & $3027^{+40}_{-42}$ & $4.74^{+0.26}_{-0.38}$ & $-11.43^{+0.25}_{-0.23}$ & $4.3^{+1.14}_{-1.33}$ & $2858^{+50}_{-100}$ & $4.74^{+0.26}_{-0.39}$ & $33.94^{+0.58}_{-0.4}$ & $5.14^{+1.88}_{-2.22}$ & $0.55^{+0.05}_{-0.05}$ & 2.6 & 4.4 \\ 
2007 Oct 26\tablenotemark{e} & 54399.55441 & 19.1 & 57 & $3001^{+33}_{-32}$ & $4.88^{+0.2}_{-0.2}$ & $-10.8^{+0.29}_{-0.22}$ & $4.11^{+1.02}_{-2.25}$ & $2962^{+50}_{-62}$ & $4.88^{+0.42}_{-0.44}$ & $33.57^{+0.35}_{-0.34}$ & $3.1^{+1.78}_{-1.57}$ & $0.56^{+0.05}_{-0.04}$ & 2.6 & 4.1 \\ 
2007 Oct 27\tablenotemark{e} & 54400.59805 & 16.5 & 57 & $2944^{+37}_{-38}$ & $5.42^{+0.05}_{-0.08}$ & $31.4^{+0.35}_{-0.25}$ & $4.43^{+1.16}_{-1.18}$ & $2978^{+76}_{-72}$ & $5.42^{+0.24}_{-0.23}$ & $-17.36^{+0.48}_{-0.55}$ & $6.93^{+1.07}_{-1.6}$ & $0.68^{+0.05}_{-0.06}$ & 2.4 & 3.8 \\ 
2007 Oct 27\tablenotemark{e} & 54400.60189 & 16.5 & 57 & $2920^{+37}_{-32}$ & $4.91^{+0.2}_{-0.18}$ & $30.89^{+0.29}_{-0.27}$ & $5.21^{+1.67}_{-1.42}$ & $2924^{+60}_{-74}$ & $4.91^{+0.43}_{-0.44}$ & $-17.64^{+0.41}_{-0.46}$ & $3.57^{+1.59}_{-1.65}$ & $0.64^{+0.06}_{-0.05}$ & 2.5 & 4.1 \\ 
2022 Mar 13 & 59651.20897 & 22.3 & 33 & $2895^{+15}_{-11}$ & $5.49^{+0.01}_{-0.01}$ & $23.84^{+0.13}_{-0.16}$ & $6.55^{+0.42}_{-0.52}$ & $2617^{+36}_{-35}$ & $5.49^{+0.02}_{-0.03}$ & $-13.04^{+0.28}_{-0.27}$ & $4.56^{+0.9}_{-1.28}$ & $0.57^{+0.01}_{-0.01}$ & 2.7 & 7.6 \\ 
2022 Mar 13 & 59651.21244 & 24.2 & 33 & $2861^{+14}_{-12}$ & $5.49^{+0.0}_{-0.01}$ & $24.43^{+0.14}_{-0.14}$ & $6.72^{+0.32}_{-0.62}$ & $2639^{+36}_{-31}$ & $5.49^{+0.01}_{-0.02}$ & $-13.83^{+0.16}_{-0.17}$ & $3.31^{+1.21}_{-1.86}$ & $0.57^{+0.01}_{-0.01}$ & 3.0 & 9.1 \\ 
2022 Mar 13 & 59651.24995 & 29.2 & 33 & $2893^{+12}_{-12}$ & $5.49^{+0.01}_{-0.01}$ & $28.22^{+0.12}_{-0.11}$ & $5.49^{+0.71}_{-0.7}$ & $2604^{+35}_{-38}$ & $5.49^{+0.03}_{-0.07}$ & $-18.75^{+0.24}_{-0.23}$ & $6.61^{+0.86}_{-0.73}$ & $0.6^{+0.01}_{-0.01}$ & 3.6 & 10.9 \\ 
2022 Mar 13 & 59651.25342 & 29.9 & 33 & $2874^{+11}_{-12}$ & $5.5^{+0.0}_{-0.01}$ & $28.5^{+0.16}_{-0.12}$ & $6.41^{+0.21}_{-0.64}$ & $2631^{+30}_{-33}$ & $5.5^{+0.01}_{-0.01}$ & $-19.82^{+0.2}_{-0.22}$ & $6.95^{+0.9}_{-0.93}$ & $0.6^{+0.02}_{-0.01}$ & 4.0 & 11.6 \\ 
2022 Mar 13 & 59651.26251 & 22.8 & 33 & $2905^{+18}_{-17}$ & $5.49^{+0.02}_{-0.01}$ & $29.07^{+0.14}_{-0.2}$ & $6.66^{+0.33}_{-0.48}$ & $2645^{+40}_{-37}$ & $5.49^{+0.03}_{-0.04}$ & $-20.52^{+0.3}_{-0.27}$ & $4.59^{+0.97}_{-2.2}$ & $0.58^{+0.02}_{-0.02}$ & 2.8 & 7.5 \\ 
2022 Mar 13 & 59651.26598 & 22.5 & 33 & $2883^{+13}_{-13}$ & $5.49^{+0.01}_{-0.01}$ & $29.31^{+0.16}_{-0.18}$ & $5.26^{+0.5}_{-0.52}$ & $2615^{+28}_{-33}$ & $5.49^{+0.02}_{-0.03}$ & $-20.68^{+0.22}_{-0.25}$ & $6.3^{+1.2}_{-1.31}$ & $0.61^{+0.02}_{-0.02}$ & 2.9 & 7.4 \\ 
2022 Jul 10 & 59770.61234 & 9.7 & 33 & $2892^{+31}_{-28}$ & $5.48^{+0.02}_{-0.04}$ & $25.69^{+0.27}_{-0.24}$ & $8.29^{+0.87}_{-0.95}$ & $2519^{+88}_{-79}$ & $5.24^{+0.19}_{-0.19}$ & $-17.25^{+0.49}_{-0.3}$ & $8.0^{+2.36}_{-2.16}$ & $0.6^{+0.05}_{-0.03}$ & 1.8 & 3.0 \\ 
2022 Jul 10 & 59770.61755 & 14.8 & 33 & $2888^{+22}_{-22}$ & $5.48^{+0.01}_{-0.02}$ & $26.59^{+0.23}_{-0.22}$ & $7.47^{+0.61}_{-0.74}$ & $2639^{+34}_{-48}$ & $5.46^{+0.03}_{-0.05}$ & $-17.96^{+0.34}_{-0.31}$ & $4.63^{+1.71}_{-1.57}$ & $0.6^{+0.03}_{-0.02}$ & 2.2 & 4.6 \\ 
2022 Jul 10 & 59770.62439 & 14.8 & 33 & $2884^{+26}_{-17}$ & $5.48^{+0.01}_{-0.03}$ & $27.06^{+0.16}_{-0.16}$ & $7.2^{+0.69}_{-0.6}$ & $2579^{+50}_{-80}$ & $5.44^{+0.05}_{-0.05}$ & $-18.76^{+0.28}_{-0.37}$ & $6.39^{+1.44}_{-1.6}$ & $0.63^{+0.02}_{-0.02}$ & 2.2 & 4.4 \\ 
2022 Jul 10 & 59770.62960 & 17.9 & 33 & $2880^{+21}_{-19}$ & $5.48^{+0.01}_{-0.03}$ & $28.28^{+0.24}_{-0.19}$ & $7.05^{+0.81}_{-0.92}$ & $2600^{+39}_{-41}$ & $5.45^{+0.03}_{-0.08}$ & $-19.28^{+0.22}_{-0.24}$ & $4.53^{+1.69}_{-1.67}$ & $0.58^{+0.02}_{-0.02}$ & 2.6 & 4.0 \\ 
2022 Oct 16 & 59868.61253 & 22.3 & 33 & $2872^{+16}_{-15}$ & $5.49^{+0.01}_{-0.01}$ & $15.41^{+0.16}_{-0.11}$ & $5.09^{+0.39}_{-0.46}$ & $2642^{+40}_{-42}$ & $5.48^{+0.02}_{-0.03}$ & $-3.97^{+0.21}_{-0.24}$ & $2.86^{+1.19}_{-1.54}$ & $0.57^{+0.03}_{-0.02}$ & 3.4 & 5.9 \\ 
2022 Oct 16 & 59868.61643 & 18.2 & 33 & $2867^{+18}_{-20}$ & $5.49^{+0.01}_{-0.02}$ & $16.07^{+0.14}_{-0.14}$ & $5.17^{+0.44}_{-0.53}$ & $2676^{+36}_{-38}$ & $5.47^{+0.02}_{-0.03}$ & $-4.96^{+0.31}_{-0.27}$ & $6.48^{+0.56}_{-0.98}$ & $0.58^{+0.03}_{-0.02}$ & 2.7 & 5.0 \\ 
2022 Oct 16 & 59868.62050 & 23.8 & 33 & $2885^{+19}_{-20}$ & $5.49^{+0.01}_{-0.02}$ & $17.07^{+0.16}_{-0.13}$ & $5.19^{+0.29}_{-0.32}$ & $2638^{+49}_{-32}$ & $5.48^{+0.01}_{-0.03}$ & $-5.74^{+0.23}_{-0.25}$ & $3.9^{+1.46}_{-1.41}$ & $0.6^{+0.03}_{-0.02}$ & 3.4 & 6.8 \\ 
2022 Oct 16 & 59868.62439 & 22.7 & 33 & $2910^{+16}_{-19}$ & $5.49^{+0.01}_{-0.01}$ & $18.16^{+0.14}_{-0.13}$ & $5.33^{+0.48}_{-0.65}$ & $2552^{+51}_{-37}$ & $5.48^{+0.01}_{-0.01}$ & $-6.52^{+0.31}_{-0.25}$ & $4.55^{+0.86}_{-1.21}$ & $0.61^{+0.02}_{-0.02}$ & 3.4 & 8.6 \\ 
2022 Oct 16 & 59868.62843 & 21.1 & 33 & $2914^{+18}_{-11}$ & $5.49^{+0.01}_{-0.01}$ & $18.47^{+0.15}_{-0.13}$ & $4.48^{+0.74}_{-0.95}$ & $2588^{+36}_{-30}$ & $5.48^{+0.01}_{-0.03}$ & $-7.48^{+0.2}_{-0.25}$ & $6.31^{+1.11}_{-0.74}$ & $0.65^{+0.02}_{-0.03}$ & 3.1 & 6.2 \\ 
2022 Oct 16 & 59868.63232 & 25.4 & 33 & $2977^{+15}_{-15}$ & $5.49^{+0.01}_{-0.01}$ & $19.75^{+0.12}_{-0.16}$ & $5.56^{+0.77}_{-0.76}$ & $2358^{+7}_{-18}$ & $5.47^{+0.02}_{-0.04}$ & $-8.13^{+0.24}_{-0.26}$ & $3.45^{+1.08}_{-1.41}$ & $0.7^{+0.04}_{-0.02}$ & 3.6 & 10.2 \\ 
2022 Oct 16 & 59868.63638 & 20.4 & 33 & $2886^{+18}_{-17}$ & $5.49^{+0.01}_{-0.02}$ & $19.96^{+0.21}_{-0.12}$ & $5.77^{+0.76}_{-0.85}$ & $2603^{+50}_{-41}$ & $5.47^{+0.02}_{-0.03}$ & $-9.51^{+0.19}_{-0.23}$ & $4.29^{+1.27}_{-1.45}$ & $0.62^{+0.02}_{-0.02}$ & 2.9 & 6.1 \\ 
2022 Oct 16 & 59868.64024 & 22.7 & 33 & $2901^{+19}_{-20}$ & $5.49^{+0.01}_{-0.01}$ & $20.45^{+0.14}_{-0.1}$ & $5.11^{+0.47}_{-0.33}$ & $2636^{+29}_{-46}$ & $5.49^{+0.01}_{-0.02}$ & $-10.45^{+0.18}_{-0.22}$ & $4.53^{+1.29}_{-1.03}$ & $0.6^{+0.02}_{-0.01}$ & 3.2 & 7.7 \\ 
2022 Oct 16 & 59868.64550 & 14.4 & 33 & $2902^{+25}_{-38}$ & $5.47^{+0.02}_{-0.05}$ & $21.57^{+0.31}_{-0.23}$ & $6.08^{+0.62}_{-0.99}$ & $2560^{+72}_{-62}$ & $5.43^{+0.05}_{-0.07}$ & $-11.7^{+0.36}_{-0.35}$ & $6.68^{+1.51}_{-1.39}$ & $0.66^{+0.02}_{-0.03}$ & 2.2 & 4.2 \\ 
2022 Oct 16 & 59868.64936 & 20.1 & 33 & $2919^{+18}_{-17}$ & $5.49^{+0.01}_{-0.02}$ & $21.91^{+0.17}_{-0.13}$ & $5.37^{+0.27}_{-0.53}$ & $2616^{+48}_{-42}$ & $5.48^{+0.01}_{-0.02}$ & $-12.33^{+0.32}_{-0.25}$ & $3.52^{+1.28}_{-1.54}$ & $0.57^{+0.02}_{-0.02}$ & 2.8 & 6.7 \\ 
2022 Oct 16\tablenotemark{e} & 59868.66123 & 7.7 & 33 & $2955^{+29}_{-42}$ & $5.46^{+0.03}_{-0.06}$ & $23.7^{+0.26}_{-0.17}$ & $3.29^{+1.03}_{-1.4}$ & $2595^{+54}_{-71}$ & $5.43^{+0.05}_{-0.1}$ & $-14.31^{+0.41}_{-0.4}$ & $5.86^{+1.96}_{-1.8}$ & $0.66^{+0.04}_{-0.04}$ & 1.5 & 2.4 \\ 
2022 Dec 2\tablenotemark{e} & 59915.47470 & 15.0 & 33 & $2930^{+28}_{-30}$ & $5.49^{+0.01}_{-0.02}$ & $6.58^{+0.27}_{-0.26}$ & $2.1^{+1.03}_{-1.27}$ & $2561^{+58}_{-94}$ & $5.49^{+0.01}_{-0.02}$ & $7.47^{+0.58}_{-0.53}$ & $9.36^{+0.46}_{-0.85}$ & $0.85^{+0.1}_{-0.17}$ & 2.4 & 3.2 \\ 
2022 Dec 2\tablenotemark{e} & 59915.47817 & 12.7 & 33 & $2893^{+31}_{-36}$ & $5.49^{+0.01}_{-0.02}$ & $6.7^{+0.57}_{-0.68}$ & $5.95^{+0.83}_{-1.08}$ & $2621^{+66}_{-53}$ & $5.48^{+0.01}_{-0.03}$ & $8.13^{+1.14}_{-0.97}$ & $8.67^{+1.06}_{-2.11}$ & $0.8^{+0.14}_{-0.2}$ & 2.0 & 2.4 \\ 
2022 Dec 2\tablenotemark{e} & 59915.48277 & 10.5 & 33 & $2911^{+50}_{-56}$ & $5.48^{+0.02}_{-0.03}$ & $8.44^{+0.7}_{-0.78}$ & $6.78^{+1.06}_{-0.84}$ & $2584^{+121}_{-113}$ & $5.47^{+0.02}_{-0.04}$ & $4.44^{+1.53}_{-1.2}$ & $8.65^{+1.1}_{-1.58}$ & $0.75^{+0.16}_{-0.16}$ & 2.0 & 2.4 \\ 
2022 Dec 2\tablenotemark{e} & 59915.48624 & 9.2 & 33 & $2820^{+51}_{-41}$ & $5.48^{+0.01}_{-0.03}$ & $10.11^{+0.62}_{-0.71}$ & $7.67^{+1.29}_{-1.42}$ & $2737^{+45}_{-68}$ & $5.47^{+0.02}_{-0.05}$ & $3.69^{+1.04}_{-1.08}$ & $8.54^{+1.09}_{-1.84}$ & $0.75^{+0.16}_{-0.17}$ & 1.6 & 2.0 \\ 
2022 Dec 2 & 59915.48637 & 15.7 & 33 & $2886^{+36}_{-34}$ & $5.48^{+0.01}_{-0.04}$ & $11.97^{+0.39}_{-0.35}$ & $3.94^{+1.21}_{-0.66}$ & $2649^{+50}_{-45}$ & $5.47^{+0.02}_{-0.03}$ & $2.08^{+0.63}_{-0.58}$ & $6.03^{+1.57}_{-1.34}$ & $0.77^{+0.1}_{-0.16}$ & 2.0 & 2.5 \\ 
2022 Dec 2 & 59915.49048 & 14.2 & 33 & $2755^{+39}_{-31}$ & $5.48^{+0.02}_{-0.02}$ & $12.14^{+0.47}_{-0.35}$ & $6.26^{+1.09}_{-0.95}$ & $2818^{+62}_{-52}$ & $5.47^{+0.02}_{-0.04}$ & $0.08^{+0.62}_{-0.51}$ & $6.06^{+1.33}_{-1.38}$ & $0.59^{+0.13}_{-0.07}$ & 2.1 & 2.6 \\ 
2022 Dec 2 & 59915.49841 & 13.3 & 33 & $2838^{+22}_{-22}$ & $5.48^{+0.01}_{-0.02}$ & $12.56^{+0.24}_{-0.24}$ & $4.34^{+0.88}_{-0.99}$ & $2671^{+57}_{-51}$ & $5.45^{+0.04}_{-0.06}$ & $-0.87^{+0.43}_{-0.39}$ & $4.44^{+1.7}_{-1.29}$ & $0.54^{+0.06}_{-0.03}$ & 2.1 & 2.6 \\ 
2022 Dec 2 & 59915.50188 & 11.7 & 33 & $2865^{+32}_{-31}$ & $5.49^{+0.01}_{-0.01}$ & $13.31^{+0.24}_{-0.22}$ & $5.24^{+0.52}_{-0.73}$ & $2619^{+63}_{-75}$ & $5.48^{+0.02}_{-0.03}$ & $-1.14^{+0.43}_{-0.38}$ & $3.83^{+1.51}_{-2.11}$ & $0.58^{+0.08}_{-0.04}$ & 1.8 & 2.3 \\ 
2022 Dec 2 & 59915.50612 & 10.8 & 33 & $2910^{+25}_{-35}$ & $5.48^{+0.02}_{-0.04}$ & $14.46^{+0.22}_{-0.26}$ & $4.86^{+0.73}_{-0.69}$ & $2605^{+67}_{-59}$ & $5.45^{+0.03}_{-0.05}$ & $-2.3^{+0.36}_{-0.34}$ & $2.68^{+1.44}_{-1.91}$ & $0.57^{+0.04}_{-0.04}$ & 1.9 & 2.4 \\ 
2022 Dec 2 & 59915.50959 & 9.2 & 33 & $2833^{+29}_{-35}$ & $5.48^{+0.02}_{-0.03}$ & $15.09^{+0.3}_{-0.31}$ & $6.74^{+0.89}_{-0.81}$ & $2689^{+66}_{-58}$ & $5.46^{+0.03}_{-0.08}$ & $-2.82^{+0.37}_{-0.4}$ & $4.47^{+1.74}_{-1.24}$ & $0.59^{+0.05}_{-0.04}$ & 1.8 & 2.3 \\ 
2022 Dec 2 & 59915.51409 & 17.5 & 33 & $2826^{+22}_{-23}$ & $5.49^{+0.01}_{-0.01}$ & $15.77^{+0.22}_{-0.21}$ & $6.05^{+0.59}_{-0.42}$ & $2724^{+35}_{-26}$ & $5.48^{+0.01}_{-0.03}$ & $-4.45^{+0.28}_{-0.2}$ & $5.19^{+0.92}_{-1.21}$ & $0.6^{+0.02}_{-0.04}$ & 2.6 & 3.8 \\ 
2022 Dec 2 & 59915.51756 & 19.2 & 33 & $2872^{+18}_{-20}$ & $5.49^{+0.01}_{-0.02}$ & $16.38^{+0.14}_{-0.11}$ & $6.12^{+0.47}_{-0.83}$ & $2660^{+46}_{-52}$ & $5.48^{+0.02}_{-0.03}$ & $-5.43^{+0.26}_{-0.26}$ & $6.23^{+0.66}_{-0.8}$ & $0.58^{+0.02}_{-0.02}$ & 2.6 & 4.4 \\ 
2022 Dec 2 & 59915.52637 & 19.8 & 33 & $2875^{+16}_{-20}$ & $5.47^{+0.02}_{-0.03}$ & $17.5^{+0.15}_{-0.16}$ & $4.93^{+0.58}_{-0.28}$ & $2637^{+43}_{-51}$ & $5.47^{+0.02}_{-0.04}$ & $-5.67^{+0.28}_{-0.25}$ & $2.48^{+1.58}_{-1.59}$ & $0.55^{+0.02}_{-0.02}$ & 2.7 & 4.9 \\ 
2022 Dec 2 & 59915.52529 & 17.3 & 33 & $2824^{+24}_{-24}$ & $5.49^{+0.01}_{-0.02}$ & $17.89^{+0.17}_{-0.14}$ & $5.53^{+0.76}_{-0.78}$ & $2745^{+31}_{-42}$ & $5.48^{+0.02}_{-0.04}$ & $-6.92^{+0.18}_{-0.27}$ & $6.3^{+0.84}_{-0.79}$ & $0.63^{+0.02}_{-0.03}$ & 2.7 & 4.6 \\ 
2022 Dec 2 & 59915.52973 & 17.7 & 33 & $2854^{+21}_{-18}$ & $5.48^{+0.01}_{-0.02}$ & $18.74^{+0.14}_{-0.21}$ & $6.17^{+0.42}_{-0.55}$ & $2722^{+39}_{-42}$ & $5.48^{+0.02}_{-0.04}$ & $-7.71^{+0.32}_{-0.29}$ & $5.92^{+0.87}_{-1.21}$ & $0.59^{+0.02}_{-0.02}$ & 2.8 & 4.9 \\ 
2022 Dec 2 & 59915.53320 & 16.6 & 33 & $2876^{+19}_{-18}$ & $5.49^{+0.01}_{-0.02}$ & $19.39^{+0.13}_{-0.11}$ & $6.32^{+0.39}_{-0.53}$ & $2627^{+49}_{-41}$ & $5.47^{+0.02}_{-0.04}$ & $-8.5^{+0.3}_{-0.25}$ & $6.37^{+1.06}_{-0.85}$ & $0.59^{+0.02}_{-0.02}$ & 2.3 & 4.5 \\ 
2022 Dec 2 & 59915.53744 & 15.8 & 33 & $2858^{+23}_{-29}$ & $5.49^{+0.01}_{-0.02}$ & $20.03^{+0.19}_{-0.13}$ & $5.48^{+0.88}_{-0.66}$ & $2693^{+44}_{-58}$ & $5.48^{+0.02}_{-0.04}$ & $-9.76^{+0.26}_{-0.27}$ & $4.19^{+1.16}_{-1.82}$ & $0.57^{+0.02}_{-0.02}$ & 2.1 & 4.4 \\ 
2022 Dec 2 & 59915.54091 & 14.5 & 33 & $2807^{+29}_{-24}$ & $5.49^{+0.01}_{-0.02}$ & $20.8^{+0.18}_{-0.18}$ & $6.66^{+0.6}_{-0.6}$ & $2705^{+53}_{-49}$ & $5.44^{+0.04}_{-0.05}$ & $-10.74^{+0.34}_{-0.33}$ & $7.95^{+1.0}_{-0.79}$ & $0.62^{+0.03}_{-0.02}$ & 2.5 & 5.0 \\ 
2022 Dec 2 & 59915.54538 & 13.3 & 33 & $2799^{+26}_{-29}$ & $5.48^{+0.01}_{-0.02}$ & $21.36^{+0.19}_{-0.18}$ & $4.43^{+0.88}_{-1.04}$ & $2750^{+51}_{-42}$ & $5.46^{+0.03}_{-0.04}$ & $-11.6^{+0.27}_{-0.3}$ & $8.31^{+0.83}_{-0.77}$ & $0.67^{+0.03}_{-0.03}$ & 2.3 & 4.3 \\ 
2022 Dec 2 & 59915.55387 & 13.6 & 33 & $2885^{+26}_{-20}$ & $5.49^{+0.01}_{-0.01}$ & $22.12^{+0.14}_{-0.18}$ & $5.51^{+0.85}_{-0.74}$ & $2608^{+51}_{-61}$ & $5.45^{+0.04}_{-0.04}$ & $-12.2^{+0.38}_{-0.32}$ & $4.54^{+1.37}_{-1.39}$ & $0.55^{+0.02}_{-0.02}$ & 2.0 & 3.9 \\ 
2022 Dec 2 & 59915.55734 & 15.0 & 33 & $2926^{+24}_{-18}$ & $5.48^{+0.01}_{-0.02}$ & $22.83^{+0.18}_{-0.22}$ & $5.39^{+0.42}_{-0.59}$ & $2643^{+43}_{-50}$ & $5.47^{+0.02}_{-0.04}$ & $-12.46^{+0.29}_{-0.29}$ & $7.35^{+1.06}_{-1.06}$ & $0.58^{+0.02}_{-0.02}$ & 2.3 & 5.1 \\ 
2023 Jan 8 & 59952.21939 & 23.6 & 33 & $2851^{+18}_{-16}$ & $5.49^{+0.01}_{-0.01}$ & $-16.3^{+0.16}_{-0.17}$ & $3.87^{+0.61}_{-0.87}$ & $2655^{+25}_{-27}$ & $5.47^{+0.02}_{-0.05}$ & $34.47^{+0.23}_{-0.23}$ & $5.24^{+1.33}_{-1.03}$ & $0.67^{+0.02}_{-0.02}$ & 2.4 & 7.4 \\ 
2023 Jan 8 & 59952.22286 & 23.6 & 33 & $2817^{+18}_{-14}$ & $5.48^{+0.01}_{-0.02}$ & $-16.53^{+0.2}_{-0.15}$ & $3.79^{+0.64}_{-0.74}$ & $2671^{+27}_{-36}$ & $5.48^{+0.02}_{-0.03}$ & $34.65^{+0.25}_{-0.25}$ & $6.08^{+0.91}_{-1.28}$ & $0.65^{+0.02}_{-0.02}$ & 2.5 & 6.5 \\ 
2023 Jan 8 & 59952.22707 & 21.3 & 33 & $2828^{+15}_{-19}$ & $5.49^{+0.01}_{-0.02}$ & $-16.76^{+0.2}_{-0.18}$ & $2.98^{+0.86}_{-1.36}$ & $2675^{+28}_{-29}$ & $5.47^{+0.02}_{-0.04}$ & $35.22^{+0.22}_{-0.22}$ & $6.46^{+0.91}_{-1.33}$ & $0.65^{+0.02}_{-0.02}$ & 2.3 & 9.1 \\ 
2023 Jan 8 & 59952.23054 & 20.8 & 33 & $2855^{+18}_{-17}$ & $5.49^{+0.01}_{-0.02}$ & $-16.95^{+0.21}_{-0.18}$ & $2.85^{+1.08}_{-1.61}$ & $2677^{+29}_{-32}$ & $5.46^{+0.03}_{-0.06}$ & $34.95^{+0.24}_{-0.31}$ & $6.63^{+1.01}_{-0.84}$ & $0.68^{+0.02}_{-0.02}$ & 2.2 & 8.1 \\ 
2023 Jan 8 & 59952.23504 & 19.9 & 33 & $2856^{+16}_{-19}$ & $5.49^{+0.01}_{-0.01}$ & $-16.92^{+0.16}_{-0.17}$ & $3.49^{+0.62}_{-1.21}$ & $2668^{+25}_{-36}$ & $5.48^{+0.01}_{-0.03}$ & $34.89^{+0.31}_{-0.31}$ & $4.58^{+1.22}_{-1.82}$ & $0.65^{+0.02}_{-0.02}$ & 2.2 & 5.5 \\ 
2023 Jan 8 & 59952.23851 & 19.3 & 33 & $2846^{+19}_{-22}$ & $5.49^{+0.01}_{-0.02}$ & $-17.11^{+0.18}_{-0.2}$ & $1.81^{+1.23}_{-0.99}$ & $2709^{+27}_{-40}$ & $5.47^{+0.02}_{-0.04}$ & $35.03^{+0.4}_{-0.4}$ & $6.22^{+1.19}_{-1.53}$ & $0.67^{+0.03}_{-0.02}$ & 2.2 & 4.8 \\ 
2023 Jan 8 & 59952.24275 & 17.2 & 33 & $2835^{+19}_{-22}$ & $5.48^{+0.01}_{-0.02}$ & $-17.0^{+0.22}_{-0.24}$ & $3.34^{+0.98}_{-1.32}$ & $2646^{+40}_{-33}$ & $5.47^{+0.02}_{-0.04}$ & $35.45^{+0.36}_{-0.42}$ & $6.25^{+1.46}_{-1.41}$ & $0.65^{+0.02}_{-0.02}$ & 2.2 & 6.3 \\ 
2023 Jan 8 & 59952.24622 & 15.9 & 33 & $2867^{+19}_{-14}$ & $5.48^{+0.01}_{-0.03}$ & $-16.78^{+0.18}_{-0.18}$ & $4.54^{+0.92}_{-1.0}$ & $2638^{+40}_{-38}$ & $5.46^{+0.03}_{-0.06}$ & $35.83^{+0.37}_{-0.37}$ & $5.71^{+1.15}_{-1.16}$ & $0.62^{+0.03}_{-0.02}$ & 2.1 & 4.5 \\ 
2023 Jan 8 & 59952.26061 & 17.1 & 33 & $2864^{+18}_{-18}$ & $5.49^{+0.01}_{-0.02}$ & $-16.43^{+0.19}_{-0.18}$ & $5.05^{+0.62}_{-0.59}$ & $2632^{+34}_{-33}$ & $5.46^{+0.03}_{-0.05}$ & $34.88^{+0.28}_{-0.34}$ & $6.27^{+0.84}_{-1.1}$ & $0.66^{+0.02}_{-0.02}$ & 2.2 & 5.5 \\ 
2023 Jan 8 & 59952.26408 & 19.6 & 33 & $2830^{+19}_{-17}$ & $5.48^{+0.01}_{-0.02}$ & $-16.45^{+0.16}_{-0.12}$ & $3.65^{+0.75}_{-1.39}$ & $2678^{+29}_{-33}$ & $5.48^{+0.01}_{-0.03}$ & $33.96^{+0.45}_{-0.46}$ & $7.5^{+0.83}_{-1.21}$ & $0.65^{+0.02}_{-0.02}$ & 2.1 & 4.7 \\ 
2023 Jan 8 & 59952.26863 & 17.6 & 33 & $2854^{+18}_{-20}$ & $5.48^{+0.01}_{-0.03}$ & $-16.07^{+0.2}_{-0.22}$ & $3.75^{+1.04}_{-1.44}$ & $2660^{+31}_{-39}$ & $5.46^{+0.03}_{-0.04}$ & $33.96^{+0.25}_{-0.4}$ & $6.53^{+1.3}_{-1.28}$ & $0.67^{+0.02}_{-0.03}$ & 2.1 & 6.4 \\ 
2023 Jan 8 & 59952.27210 & 17.4 & 33 & $2848^{+14}_{-16}$ & $5.49^{+0.01}_{-0.01}$ & $-15.84^{+0.18}_{-0.18}$ & $3.02^{+1.02}_{-1.16}$ & $2689^{+19}_{-26}$ & $5.48^{+0.02}_{-0.03}$ & $33.97^{+0.25}_{-0.31}$ & $6.13^{+0.98}_{-1.19}$ & $0.64^{+0.02}_{-0.02}$ & 2.3 & 5.8 \\ 
2023 Jan 8 & 59952.27630 & 20.0 & 33 & $2835^{+22}_{-15}$ & $5.49^{+0.01}_{-0.01}$ & $-15.86^{+0.27}_{-0.21}$ & $3.75^{+1.26}_{-0.82}$ & $2682^{+25}_{-39}$ & $5.48^{+0.02}_{-0.02}$ & $33.6^{+0.25}_{-0.25}$ & $5.01^{+1.24}_{-1.54}$ & $0.65^{+0.02}_{-0.02}$ & 3.3 & 8.4 \\ 
2023 Jan 8 & 59952.27977 & 24.3 & 33 & $2858^{+13}_{-16}$ & $5.49^{+0.01}_{-0.01}$ & $-15.57^{+0.14}_{-0.14}$ & $3.86^{+0.62}_{-1.34}$ & $2660^{+28}_{-34}$ & $5.46^{+0.03}_{-0.04}$ & $33.6^{+0.25}_{-0.32}$ & $7.71^{+0.73}_{-0.9}$ & $0.68^{+0.02}_{-0.02}$ & 2.6 & 7.4 \\ 
2023 Jan 8 & 59952.28447 & 23.9 & 33 & $2891^{+27}_{-19}$ & $5.49^{+0.01}_{-0.02}$ & $-15.88^{+0.26}_{-0.23}$ & $3.16^{+1.08}_{-1.41}$ & $2622^{+49}_{-39}$ & $5.45^{+0.03}_{-0.07}$ & $32.28^{+0.3}_{-0.34}$ & $5.01^{+1.57}_{-1.86}$ & $0.66^{+0.02}_{-0.03}$ & 3.1 & 10.3 \\ 
2023 Jan 8 & 59952.28794 & 23.8 & 33 & $2870^{+18}_{-16}$ & $5.49^{+0.01}_{-0.02}$ & $-15.4^{+0.17}_{-0.22}$ & $4.13^{+1.04}_{-0.87}$ & $2638^{+34}_{-34}$ & $5.48^{+0.02}_{-0.03}$ & $32.05^{+0.34}_{-0.46}$ & $3.97^{+1.55}_{-1.82}$ & $0.66^{+0.02}_{-0.02}$ & 2.3 & 5.5 \\ 
2023 Jan 8 & 59952.29219 & 18.5 & 33 & $2800^{+55}_{-47}$ & $5.3^{+0.1}_{-0.1}$ & $-15.85^{+0.55}_{-0.57}$ & $1.07^{+1.36}_{-0.78}$ & $2749^{+67}_{-89}$ & $5.39^{+0.07}_{-0.17}$ & $29.77^{+0.35}_{-0.4}$ & $4.13^{+2.25}_{-2.4}$ & $0.67^{+0.05}_{-0.05}$ & 2.1 & 3.0 \\ 
2023 Jan 8 & 59952.29566 & 11.5 & 33 & $2835^{+45}_{-45}$ & $5.4^{+0.08}_{-0.13}$ & $-14.82^{+0.59}_{-0.72}$ & $4.96^{+1.88}_{-1.74}$ & $2614^{+64}_{-86}$ & $5.3^{+0.12}_{-0.18}$ & $31.48^{+0.83}_{-0.83}$ & $2.46^{+2.35}_{-1.82}$ & $0.71^{+0.05}_{-0.05}$ & 1.7 & 2.0 \\ 
2023 Jan 8 & 59952.30045 & 5.9 & 33 & $2910^{+45}_{-40}$ & $5.45^{+0.04}_{-0.1}$ & $-14.06^{+0.32}_{-0.28}$ & $2.28^{+1.83}_{-1.2}$ & $2585^{+71}_{-85}$ & $5.31^{+0.15}_{-0.13}$ & $31.98^{+0.58}_{-0.58}$ & $3.81^{+2.44}_{-2.82}$ & $0.7^{+0.04}_{-0.04}$ & 1.8 & 2.7 \\ 
2023 Jan 8 & 59952.30392 & 8.7 & 33 & $2889^{+19}_{-21}$ & $5.48^{+0.02}_{-0.03}$ & $-13.65^{+0.19}_{-0.17}$ & $3.93^{+1.26}_{-1.25}$ & $2569^{+11}_{-23}$ & $5.47^{+0.02}_{-0.03}$ & $30.79^{+0.27}_{-0.27}$ & $1.97^{+1.56}_{-1.09}$ & $0.63^{+0.02}_{-0.02}$ & 2.9 & 5.7 \\ 
2023 Jan 8 & 59952.30816 & 19.2 & 33 & $2845^{+25}_{-39}$ & $5.48^{+0.02}_{-0.03}$ & $-12.76^{+0.35}_{-0.25}$ & $4.1^{+1.1}_{-1.25}$ & $2687^{+42}_{-44}$ & $5.42^{+0.05}_{-0.11}$ & $31.34^{+0.45}_{-0.77}$ & $2.88^{+2.05}_{-1.96}$ & $0.62^{+0.04}_{-0.03}$ & 2.3 & 3.7 \\ 
2023 Jan 8 & 59952.31163 & 13.5 & 33 & $2908^{+30}_{-31}$ & $5.46^{+0.03}_{-0.06}$ & $-12.36^{+0.26}_{-0.27}$ & $4.82^{+0.88}_{-1.46}$ & $2582^{+63}_{-50}$ & $5.44^{+0.04}_{-0.05}$ & $29.76^{+0.44}_{-0.46}$ & $2.58^{+2.33}_{-1.79}$ & $0.63^{+0.02}_{-0.02}$ & 2.5 & 4.5 \\ 
2023 Jan 8 & 59952.31611 & 16.3 & 33 & $2876^{+16}_{-23}$ & $5.48^{+0.01}_{-0.02}$ & $-12.35^{+0.25}_{-0.23}$ & $2.43^{+1.24}_{-1.44}$ & $2646^{+41}_{-48}$ & $5.46^{+0.02}_{-0.08}$ & $28.59^{+0.3}_{-0.34}$ & $4.4^{+1.7}_{-2.22}$ & $0.65^{+0.02}_{-0.02}$ & 3.0 & 6.2 \\ 
2023 Jan 8 & 59952.31958 & 20.1 & 33 & $2864^{+22}_{-24}$ & $5.49^{+0.01}_{-0.02}$ & $-11.55^{+0.32}_{-0.19}$ & $2.05^{+1.16}_{-1.17}$ & $2665^{+43}_{-44}$ & $5.46^{+0.03}_{-0.05}$ & $29.02^{+0.38}_{-0.38}$ & $3.84^{+1.66}_{-1.71}$ & $0.69^{+0.02}_{-0.02}$ & 2.7 & 5.8 \\ 
2023 Jan 8 & 59952.32379 & 17.4 & 33 & $2797^{+24}_{-21}$ & $5.49^{+0.01}_{-0.02}$ & $-11.99^{+0.15}_{-0.17}$ & $1.97^{+1.04}_{-1.12}$ & $2776^{+20}_{-8}$ & $5.48^{+0.02}_{-0.03}$ & $27.35^{+0.34}_{-0.34}$ & $6.82^{+1.27}_{-1.37}$ & $0.72^{+0.04}_{-0.03}$ & 3.0 & 6.2 \\ 
2023 Jan 8 & 59952.32726 & 18.0 & 33 & $2881^{+16}_{-19}$ & $5.48^{+0.02}_{-0.03}$ & $-10.64^{+0.17}_{-0.26}$ & $2.32^{+0.85}_{-1.22}$ & $2640^{+32}_{-37}$ & $5.45^{+0.03}_{-0.06}$ & $27.07^{+0.54}_{-0.54}$ & $3.9^{+1.44}_{-1.68}$ & $0.66^{+0.02}_{-0.02}$ & 2.7 & 7.5 \\ 
2023 Jan 8 & 59952.33181 & 22.3 & 33 & $2907^{+21}_{-27}$ & $5.47^{+0.02}_{-0.02}$ & $-9.68^{+0.16}_{-0.21}$ & $3.03^{+1.08}_{-1.04}$ & $2629^{+47}_{-51}$ & $5.46^{+0.02}_{-0.04}$ & $26.35^{+0.37}_{-0.37}$ & $3.63^{+1.89}_{-1.95}$ & $0.64^{+0.02}_{-0.02}$ & 3.3 & 8.1 \\ 
2023 Jan 8 & 59952.33527 & 24.7 & 33 & $2878^{+15}_{-13}$ & $5.49^{+0.01}_{-0.01}$ & $-9.16^{+0.18}_{-0.18}$ & $2.17^{+0.83}_{-1.38}$ & $2604^{+32}_{-40}$ & $5.47^{+0.02}_{-0.04}$ & $26.13^{+0.32}_{-0.32}$ & $3.84^{+1.67}_{-1.8}$ & $0.66^{+0.02}_{-0.02}$ & 2.5 & 5.9 \\ 
2023 Jan 8 & 59952.33951 & 19.1 & 33 & $2878^{+29}_{-21}$ & $5.48^{+0.02}_{-0.02}$ & $-9.19^{+0.3}_{-0.34}$ & $2.23^{+0.98}_{-1.22}$ & $2592^{+59}_{-56}$ & $5.45^{+0.03}_{-0.05}$ & $24.84^{+0.31}_{-0.37}$ & $5.54^{+1.75}_{-2.18}$ & $0.68^{+0.02}_{-0.02}$ & 3.1 & 6.1 \\ 
2023 Jan 8 & 59952.34298 & 20.0 & 33 & $2883^{+16}_{-22}$ & $5.48^{+0.01}_{-0.02}$ & $-8.22^{+0.25}_{-0.42}$ & $4.23^{+1.03}_{-0.72}$ & $2572^{+48}_{-50}$ & $5.47^{+0.02}_{-0.03}$ & $24.13^{+0.59}_{-0.59}$ & $7.38^{+1.21}_{-1.07}$ & $0.69^{+0.02}_{-0.02}$ & 3.1 & 6.7 \\ 
2023 Jan 8 & 59952.34754 & 23.2 & 33 & $2890^{+27}_{-18}$ & $5.49^{+0.01}_{-0.02}$ & $-7.44^{+0.18}_{-0.19}$ & $3.84^{+0.97}_{-0.63}$ & $2580^{+46}_{-48}$ & $5.47^{+0.02}_{-0.03}$ & $23.92^{+0.41}_{-0.41}$ & $7.57^{+1.42}_{-1.46}$ & $0.68^{+0.02}_{-0.03}$ & 2.9 & 6.2 \\ 
2023 Jan 8 & 59952.35101 & 23.1 & 33 & $2864^{+28}_{-37}$ & $5.48^{+0.01}_{-0.03}$ & $-6.32^{+0.18}_{-0.21}$ & $1.02^{+0.97}_{-0.61}$ & $2622^{+67}_{-56}$ & $5.48^{+0.02}_{-0.04}$ & $23.72^{+0.44}_{-0.44}$ & $7.0^{+1.42}_{-1.54}$ & $0.72^{+0.03}_{-0.03}$ & 3.0 & 5.3 \\ 
2023 Jan 8 & 59952.35525 & 20.1 & 33 & $2905^{+25}_{-25}$ & $5.49^{+0.01}_{-0.02}$ & $-6.31^{+0.22}_{-0.31}$ & $3.68^{+1.36}_{-0.87}$ & $2557^{+54}_{-51}$ & $5.47^{+0.02}_{-0.05}$ & $21.59^{+0.35}_{-0.36}$ & $6.93^{+2.18}_{-1.94}$ & $0.7^{+0.03}_{-0.03}$ & 2.7 & 4.4 \\ 
2023 Jan 8 & 59952.35872 & 19.0 & 33 & $2889^{+15}_{-19}$ & $5.49^{+0.01}_{-0.01}$ & $-5.89^{+0.19}_{-0.26}$ & $4.31^{+0.95}_{-0.75}$ & $2528^{+48}_{-38}$ & $5.48^{+0.01}_{-0.03}$ & $20.67^{+0.29}_{-0.34}$ & $8.44^{+1.28}_{-1.23}$ & $0.73^{+0.03}_{-0.02}$ & 2.4 & 4.3 \\ 
2023 Jan 8 & 59952.36321 & 20.4 & 33 & $2882^{+29}_{-28}$ & $5.49^{+0.01}_{-0.02}$ & $-5.01^{+0.29}_{-0.42}$ & $3.03^{+1.29}_{-1.58}$ & $2619^{+57}_{-66}$ & $5.49^{+0.01}_{-0.01}$ & $19.64^{+0.35}_{-0.35}$ & $8.8^{+1.25}_{-1.23}$ & $0.73^{+0.03}_{-0.03}$ & 2.5 & 4.7 \\ 
2023 Jan 8 & 59952.36668 & 23.2 & 33 & $2897^{+21}_{-13}$ & $5.49^{+0.01}_{-0.01}$ & $-3.98^{+0.17}_{-0.19}$ & $1.51^{+1.09}_{-0.96}$ & $2545^{+33}_{-32}$ & $5.46^{+0.03}_{-0.04}$ & $18.89^{+0.33}_{-0.33}$ & $7.16^{+1.32}_{-1.39}$ & $0.76^{+0.04}_{-0.03}$ & 3.0 & 4.5 \\ 
2023 Jan 8 & 59952.37092 & 21.0 & 33 & $2914^{+18}_{-18}$ & $5.49^{+0.01}_{-0.01}$ & $-3.27^{+0.19}_{-0.19}$ & $1.9^{+1.16}_{-1.16}$ & $2550^{+36}_{-42}$ & $5.46^{+0.02}_{-0.05}$ & $18.09^{+0.3}_{-0.3}$ & $6.79^{+1.02}_{-1.18}$ & $0.73^{+0.03}_{-0.02}$ & 3.0 & 4.1 \\ 
2023 Jan 8 & 59952.37439 & 21.0 & 33 & $2936^{+29}_{-34}$ & $5.49^{+0.01}_{-0.01}$ & $-2.93^{+0.25}_{-0.29}$ & $2.45^{+1.45}_{-1.37}$ & $2527^{+42}_{-50}$ & $5.46^{+0.03}_{-0.04}$ & $15.95^{+0.54}_{-0.7}$ & $10.36^{+1.65}_{-1.44}$ & $0.89^{+0.06}_{-0.06}$ & 2.4 & 3.0 \\ 
2023 Jan 8 & 59952.37893 & 18.8 & 33 & $2943^{+25}_{-34}$ & $5.48^{+0.01}_{-0.02}$ & $-2.24^{+0.28}_{-0.2}$ & $3.16^{+1.04}_{-1.22}$ & $2515^{+43}_{-68}$ & $5.46^{+0.03}_{-0.05}$ & $15.48^{+0.41}_{-0.53}$ & $10.69^{+1.31}_{-1.32}$ & $0.94^{+0.04}_{-0.05}$ & 2.2 & 2.6 \\ 
2023 Jan 8 & 59952.38240 & 16.8 & 33 & $2923^{+25}_{-21}$ & $5.49^{+0.01}_{-0.01}$ & $-1.21^{+0.24}_{-0.24}$ & $1.5^{+1.13}_{-1.09}$ & $2535^{+38}_{-42}$ & $5.4^{+0.06}_{-0.09}$ & $14.78^{+0.4}_{-0.41}$ & $6.99^{+1.62}_{-1.7}$ & $0.77^{+0.07}_{-0.05}$ & 2.5 & 2.9 \\ 
2023 Jan 8 & 59952.38664 & 18.9 & 33 & $2947^{+34}_{-22}$ & $5.49^{+0.01}_{-0.01}$ & $-0.05^{+0.22}_{-0.21}$ & $1.2^{+1.33}_{-0.88}$ & $2549^{+53}_{-48}$ & $5.46^{+0.03}_{-0.04}$ & $13.97^{+0.52}_{-0.52}$ & $6.97^{+1.9}_{-1.28}$ & $0.87^{+0.07}_{-0.07}$ & 2.3 & 2.9 \\ 
2023 Jan 8 & 59952.39011 & 19.5 & 33 & $2888^{+26}_{-17}$ & $5.49^{+0.01}_{-0.01}$ & $1.23^{+0.53}_{-0.38}$ & $6.46^{+1.16}_{-0.95}$ & $2433^{+51}_{-82}$ & $5.43^{+0.05}_{-0.1}$ & $14.56^{+0.75}_{-0.95}$ & $4.38^{+1.77}_{-2.5}$ & $0.59^{+0.08}_{-0.1}$ & 2.4 & 3.0 \\ 
2023 Jan 8 & 59952.39459 & 21.8 & 33 & $2908^{+29}_{-30}$ & $5.49^{+0.01}_{-0.01}$ & $1.44^{+0.42}_{-0.55}$ & $5.46^{+1.28}_{-1.47}$ & $2488^{+63}_{-68}$ & $5.48^{+0.01}_{-0.02}$ & $13.04^{+0.98}_{-0.98}$ & $2.75^{+2.68}_{-1.94}$ & $0.65^{+0.11}_{-0.11}$ & 2.1 & 2.6 \\ 
2023 Jan 8 & 59952.39806 & 19.1 & 33 & $2939^{+19}_{-32}$ & $5.49^{+0.01}_{-0.02}$ & $1.07^{+0.33}_{-0.27}$ & $1.82^{+1.17}_{-1.11}$ & $2566^{+39}_{-58}$ & $5.47^{+0.02}_{-0.05}$ & $10.78^{+0.33}_{-0.37}$ & $1.69^{+1.97}_{-1.22}$ & $0.92^{+0.05}_{-0.1}$ & 2.7 & 2.7 \\ 
2023 Jan 8 & 59952.40230 & 16.9 & 33 & $2890^{+49}_{-33}$ & $5.49^{+0.01}_{-0.01}$ & $3.34^{+0.59}_{-0.57}$ & $4.73^{+1.0}_{-1.28}$ & $2488^{+93}_{-88}$ & $5.47^{+0.02}_{-0.07}$ & $11.41^{+1.26}_{-1.4}$ & $2.93^{+2.8}_{-1.92}$ & $0.65^{+0.26}_{-0.21}$ & 2.9 & 3.0 \\ 
2023 Jan 8 & 59952.40577 & 19.5 & 33 & $2889^{+28}_{-18}$ & $5.49^{+0.01}_{-0.01}$ & $4.28^{+0.3}_{-0.39}$ & $3.35^{+1.06}_{-1.17}$ & $2495^{+85}_{-74}$ & $5.47^{+0.02}_{-0.03}$ & $9.76^{+1.11}_{-1.11}$ & $10.75^{+1.51}_{-2.46}$ & $0.64^{+0.18}_{-0.1}$ & 2.4 & 2.8 \\ 
2023 Jan 8 & 59952.41024 & 19.1 & 33 & $2890^{+31}_{-26}$ & $5.49^{+0.01}_{-0.02}$ & $4.01^{+0.65}_{-0.44}$ & $2.41^{+1.35}_{-1.77}$ & $2586^{+35}_{-49}$ & $5.48^{+0.02}_{-0.04}$ & $9.01^{+0.86}_{-0.86}$ & $7.73^{+3.01}_{-1.72}$ & $0.73^{+0.15}_{-0.14}$ & 2.1 & 2.5 \\ 
2023 Jan 8 & 59952.41371 & 16.6 & 33 & $2887^{+25}_{-29}$ & $5.49^{+0.0}_{-0.02}$ & $5.14^{+0.67}_{-0.74}$ & $3.3^{+1.88}_{-1.79}$ & $2625^{+46}_{-38}$ & $5.49^{+0.01}_{-0.02}$ & $7.98^{+1.02}_{-1.02}$ & $4.53^{+2.94}_{-2.84}$ & $0.83^{+0.09}_{-0.16}$ & 3.4 & 3.4 \\ 
2023 Jan 8 & 59952.41793 & 19.7 & 33 & $2875^{+30}_{-16}$ & $5.49^{+0.01}_{-0.01}$ & $5.63^{+0.51}_{-0.7}$ & $1.72^{+1.3}_{-1.25}$ & $2608^{+30}_{-24}$ & $5.49^{+0.01}_{-0.02}$ & $7.55^{+0.94}_{-0.95}$ & $6.41^{+1.8}_{-2.65}$ & $0.72^{+0.2}_{-0.11}$ & 3.2 & 3.8 \\ 
2023 Jan 8 & 59952.42140 & 20.8 & 33 & $2847^{+29}_{-35}$ & $5.49^{+0.01}_{-0.02}$ & $5.84^{+0.32}_{-0.61}$ & $2.2^{+1.28}_{-1.2}$ & $2627^{+54}_{-50}$ & $5.48^{+0.02}_{-0.03}$ & $7.8^{+0.76}_{-0.79}$ & $10.06^{+1.6}_{-2.07}$ & $0.72^{+0.15}_{-0.15}$ & 2.1 & 2.5 \\ 
2023 Jan 8 & 59952.42584 & 14.7 & 33 & $2869^{+28}_{-34}$ & $5.49^{+0.01}_{-0.02}$ & $6.16^{+0.58}_{-0.77}$ & $4.05^{+1.46}_{-2.08}$ & $2673^{+46}_{-53}$ & $5.49^{+0.01}_{-0.02}$ & $6.9^{+0.78}_{-0.98}$ & $6.53^{+2.08}_{-3.17}$ & $0.77^{+0.15}_{-0.15}$ & 2.4 & 2.8 \\ 
2023 Jan 8 & 59952.42931 & 18.7 & 33 & $2881^{+27}_{-21}$ & $5.49^{+0.01}_{-0.02}$ & $6.76^{+0.52}_{-0.43}$ & $2.17^{+1.77}_{-1.37}$ & $2614^{+50}_{-34}$ & $5.49^{+0.01}_{-0.02}$ & $6.21^{+0.85}_{-0.85}$ & $7.89^{+1.55}_{-1.73}$ & $0.87^{+0.1}_{-0.16}$ & 2.6 & 3.1 \\ 
2023 Jan 8\tablenotemark{e} & 59952.43355 & 19.1 & 33 & $2860^{+34}_{-34}$ & $5.49^{+0.01}_{-0.02}$ & $7.06^{+0.59}_{-0.45}$ & $5.13^{+1.73}_{-1.87}$ & $2657^{+57}_{-51}$ & $5.48^{+0.01}_{-0.03}$ & $6.45^{+1.03}_{-1.03}$ & $10.31^{+1.64}_{-1.92}$ & $0.82^{+0.11}_{-0.16}$ & 2.4 & 2.9 \\ 
2023 Jan 8\tablenotemark{e} & 59952.43702 & 18.2 & 33 & $2904^{+30}_{-33}$ & $5.49^{+0.01}_{-0.01}$ & $7.29^{+0.55}_{-0.85}$ & $7.35^{+0.96}_{-1.46}$ & $2626^{+45}_{-47}$ & $5.49^{+0.01}_{-0.01}$ & $6.49^{+0.74}_{-1.23}$ & $10.18^{+1.8}_{-2.33}$ & $0.84^{+0.11}_{-0.11}$ & 2.2 & 2.7 \\ 
2023 Jan 8\tablenotemark{e} & 59952.44147 & 19.1 & 33 & $2865^{+30}_{-29}$ & $5.49^{+0.01}_{-0.01}$ & $7.47^{+0.75}_{-0.95}$ & $9.73^{+1.54}_{-1.05}$ & $2633^{+52}_{-49}$ & $5.49^{+0.01}_{-0.02}$ & $6.72^{+1.02}_{-1.43}$ & $12.76^{+1.61}_{-2.07}$ & $0.84^{+0.12}_{-0.16}$ & 1.9 & 2.3 \\ 
2023 Jan 8 & 59952.44954 & 17.6 & 33 & $2866^{+19}_{-22}$ & $5.48^{+0.02}_{-0.03}$ & $11.52^{+0.35}_{-0.28}$ & $5.72^{+1.03}_{-1.13}$ & $2534^{+78}_{-86}$ & $5.41^{+0.06}_{-0.1}$ & $0.14^{+0.38}_{-0.46}$ & $1.5^{+1.68}_{-1.06}$ & $0.53^{+0.04}_{-0.02}$ & 2.5 & 2.5 \\ 
2023 Jan 8 & 59952.45301 & 16.7 & 33 & $2827^{+25}_{-23}$ & $5.49^{+0.01}_{-0.02}$ & $12.92^{+0.33}_{-0.22}$ & $2.25^{+1.19}_{-1.21}$ & $2697^{+38}_{-42}$ & $5.49^{+0.01}_{-0.02}$ & $-0.4^{+0.4}_{-0.47}$ & $2.7^{+2.27}_{-1.46}$ & $0.68^{+0.09}_{-0.06}$ & 2.0 & 2.5 \\}
\hline
\replaced{Average\tablenotemark{e} & \nodata & \nodata & \nodata & $2877 \pm 41$ & $5.485 \pm 0.008$ & \nodata & $5.7 \pm 1.0$ & $2634 \pm 75$ & $5.46 \pm 0.04$ & \nodata & $5.3 \pm 1.5$ & $0.60 \pm 0.04$ & \nodata & \nodata}
{Average & \nodata & \nodata & \nodata & $2879 \pm 37$ & $5.49 \pm 0.02$ & \nodata & $4.2 \pm 1.7$ & $2627 \pm 71$ & $5.47 \pm 0.04$ & \nodata & $5.9 \pm 2.0$ & $0.65 \pm 0.08$ & \nodata & \nodata}
\enddata 
\tablenotetext{a}{Modified Julian Dates\added{ with the Barycentric Dynamical Time correction included following \cite{Eastman:2010aa}}}
\tablenotetext{b}{Flux scale for the secondary component}
\tablenotetext{c}{Reduced $\chi^2$} 
\tablenotetext{d}{Solution for the single star model}
\tablenotetext{e}{\replaced{Averaged over all observations with the exception of all order 57 data from 2007 Oct, the last epoch in 2022 Oct 16 (affected by twilight), and the first four epochs of 2022 Dec 2 (spectral components unresolved).}{Epochs not included in parameter averages or orbit fits.}}
\end{deluxetable} 
\end{longrotatetable}

\clearpage

\bibliography{main}{}
\bibliographystyle{aasjournal}

\end{document}